\documentclass[11pt,a4paper]{article}
\usepackage{jheppub}
\bibstyle{JHEP}

\usepackage[english]{babel}
\usepackage[utf8]{inputenc}
\usepackage[T1]{fontenc}
\usepackage{amsfonts}
\usepackage{amsmath}       
\usepackage{amsfonts}          
\usepackage{amssymb}
\usepackage{mathtools}
\usepackage{comment}
\usepackage{tikz}
\usetikzlibrary{decorations.markings}
\usepackage{physics}
\usepackage{dsfont}
\usepackage{tabu}
\usepackage{tikz}
\usetikzlibrary{arrows.meta}
\usepackage{cleveref}
\usepackage{placeins}
\usepackage{subcaption}

\title{Density kernel and temporal quantum information probes in double-scaled SYK}

\author[a]{Johanna Erdmenger,}
\author[a]{Maximilian Gaschler,}
\author[a,b]{Jonathan Karl,}
\author[c]{Arnab Kundu,}
\author[a]{and Georg Setzer}
\date{}
\affiliation[a]{Institute for Theoretical Physics and Astrophysics and Würzburg-Dresden Cluster of Excellence
ctd.qmat, Julius-Maximilians-Universit\"at W\"urzburg, Am Hubland, 97074 Würzburg, Germany}
\affiliation[b]{Department of Physics, Ben-Gurion University of the Negev,
David Ben-Gurion Boulevard 1, Beer Sheva 84105, Israel}
\affiliation[c]{Theory Division, Saha Institute of Nuclear Physics, A CI of Homi Bhabha National Institute, 1/AF, Bidhannagar, Kolkata 700064, India}

\emailAdd{maximilian.gaschler@uni-wuerzburg.de}
\emailAdd{jonathan.karl@uni-wuerzburg.de}
\emailAdd{arnab.kundu@saha.ac.in}

\abstract{Entanglement is conventionally defined through a Hilbert space bipartition on a fixed time slice. Recent work has generalized this notion to multiple times through a non-Hermitian density kernel constructed from temporal correlation functions, encoding causal response, temporal non-classicality and quantum dynamics. We construct this kernel in the double-scaled Sachdev–Ye–Kitaev (DSSYK) model using its exact chord Hilbert space formulation and study the resulting temporal information diagnostics. The direct sum structure of chord space naturally introduces a coarse-graining under which coherence between different chord sectors is removed. We show that the corresponding kernel measures quantify how information is dynamically transferred across subregions in the chord Hilbert space. We find that decreasing $\lambda$ enhances the propagation of temporal correlation across chord sectors, as well as the loss of memory within a fixed sector. This accelerated temporal dynamics in the small $\lambda$ limit coincides with the emergence of a semiclassical gravity description. Depending on the chord space partition, the probes exhibit either saturation, a growth followed by a decay, or temporary memory loss. The characteristic propagation time is nicely captured by a wavefront picture in Krylov space, with position-dependent hopping parameters that are determined by the DSSYK Lanczos coefficients. Finally, given the quantum sine-dilaton gravity description of DSSYK and its connections to de Sitter holography, our results provide a controlled setting for probing temporal quantum correlations across quantum and semiclassical gravitational regimes, and motivate a geometric interpretation of the density kernel. }

\begin{document}
\maketitle

\section{Introduction}

Quantum entanglement plays a very crucial role in the modern understanding of salient aspects of quantum gravity, such as Holography. The best understood example of this is the class of AdS/CFT correspondences \cite{Maldacena:1997,Witten:1998qj,Gubser:1998bc}, in which the entanglement entropy of a reduced density matrix corresponding to a spatial subregion is obtained by a geometric area extremization in the asymptotically AdS background \cite{Ryu:2006bv,Ryu:2006ef}. While this has been extremely successful, several intrinsic dynamical aspects of a quantum system are captured by correlation functions with operators inserted at different time-slices. This sheds light on the nature of causal response, operator growth, the physics of scrambling, and spectral correlations, among other key features \cite{Maldacena:2015waa,Cotler:2016fpe,Saad:2018bqo,Parker:2018yvk,Saad:2019lba}. 

By construction, the density matrix is defined on a Cauchy slice. It is therefore natural to seek a generalization of the density operator in which unequal time correlation functions are a part of the input data. This generalization comes at the cost of Hermiticity and positivity of the density operator.
In recent years, there have been several approaches that point in this direction. For example, it is straightforward to analytically continue the standard entanglement entropy for a given spatial subregion into a time-domain. This is naturally connected with the notion of pseudoentropy and non-Hermitian transition matrices, and has been especially explored in the context of Holography \cite{Doi_2023}. Moreover, these ideas emerge naturally in the context of de Sitter Holography, where timelike entanglement in the CFT and pseudoentropy in dS/CFT suggest that quantum entanglement across time is a more fundamental ingredient of the underlying description \cite{Doi:2022iyj}.  

Motivated by the above, it is desirable to provide an independent definition of timelike entanglement which is not just an analytic continuation of a known spatial entanglement. It turns out that such a natural definition indeed exists in which the multi-time correlation function is directly used. The corresponding correlation function of operators at different time-slices can be subsequently encoded in a generalized spacetime density kernel (GSDK), as explicitly proposed in \cite{Milekhin:2025ycm} and further studied in \cite{Das:2025fcd,Das:2026ifj}. As expected, the density kernel is no longer Hermitian or positive. Both of these departures have physical origins. Lack of Hermiticity encodes the inequivalence of time-ordering between operators \cite{Milekhin:2025ycm}, while the lack of positivity quantitatively captures non-classicality of the quantum system \cite{Das:2026ifj}. Note that, in this description which is inherently Lorentzian, there is no need to invoke an analytic continuation. It is also important to note that the density kernel framework can be shown to capture {\it e.g.}~both early-time and late-time dynamical quantum chaotic features of Lyapunov growth and spectral correlations, respectively \cite{Das:2025fcd,Das:2026ifj}. In this respect, the density kernel approach provides us with a robust framework in which temporal entanglement can be addressed for a generic quantum mechanical system.   

In particular, this framework provides an explicit connection between quantum information and quantum chaos. The kernel obtained from a $(2n)$-point correlation function with the maximal non-trivial time-ordering, upon a suitable Haar-averaging, yields the $(2n)$-th moment of the spectral form factor of the system \cite{Das:2025fcd}. More recently, several density kernel derived probes, {\it e.g.}~the Frobenius norm, Tsallis entropy, imagitivity, and negativity are explicitly shown to distinguish between a chaotic, an integrable and a multifractal phase in the Rosenzweig--Porter model \cite{Das:2026ifj}. These results provide a strong indication that the density kernel contains the information of temporal quantum correlations, non-classical character and the underlying quantum dynamics, all at once. 

Given these, it is natural to probe aspects of quantum gravity from this perspective. This is particularly interesting for de Sitter space, in which the observer clock and causal access is a fundamental part of the description \cite{Chandrasekaran:2022cip}. Correspondingly, the available information is associated with the causal diamond of the observer and its evolution in proper time. Temporal quantum correlations, therefore, become a natural object of interest for de Sitter. Here, we take the first step towards studying such temporal correlations using the density kernel.

A particularly promising setup for studying de Sitter space from a holographic perspective is realized in the double-scaled SYK (DSSYK) model \cite{Berkooz:2018qkz,Berkooz:2018jqr}. The SYK model \cite{SachdevYe:93,Kitaev:2015a} already combines strong interaction, a soluble large-$N$ regime, and maximal chaos in the IR and is holographically related to nearly-${\rm AdS}_2$ dynamics \cite{Polchinski2016,Maldacena:2016hyu,Maldacena:2016upp}. In the double-scaling limit, the DSSYK model acquires an exact description in terms of chord diagrams and an auxiliary q-deformed oscillator \cite{Berkooz:2018qkz,Berkooz:2018jqr}. As a result, several observables become exactly solvable in a highly controlled framework. Furthermore, in a series of works \cite{Blommaert:2023opb,Blommaert:2024ymv,Blommaert:2024whf,Blommaert:2025avl}, a precise duality between this model and sine dilaton gravity, which exhibits features of both de Sitter and Anti-de Sitter gravity in different regimes, has been shown. In addition, the DSSYK model exhibits features that hint at a relation to three-dimensional de Sitter space \cite{Susskind:2022bia,Lin:2022nss,Rahman:2022jsf,Narovlansky:2023lfz,Verlinde:2024znh,Verlinde:2024zrh,Gaiotto:2024kze,Tietto:2025oxn,Narovlansky:2025tpb,Blommaert:2025eps}: A particularly constrained pair of DSSYK models has been proposed as the microscopic dual to de Sitter physics \cite{Narovlansky:2023lfz}. From this perspective, temporal quantum information observables in the DSSYK model are not only interesting for understanding the quantum mechanical model, it also provides us with a controlled framework in which the emerging (quantum) gravitational aspects of de Sitter can be studied. Remarkably, the DSSYK model is equipped with a coupling, which we denote as $\lambda$, that allows us to interpolate between the true DSSYK spectrum and the in the Lyapunov sense maximally chaotic semiclassical gravity limit. 

\begin{figure}
    \centering
    \includegraphics[width=0.6\linewidth]{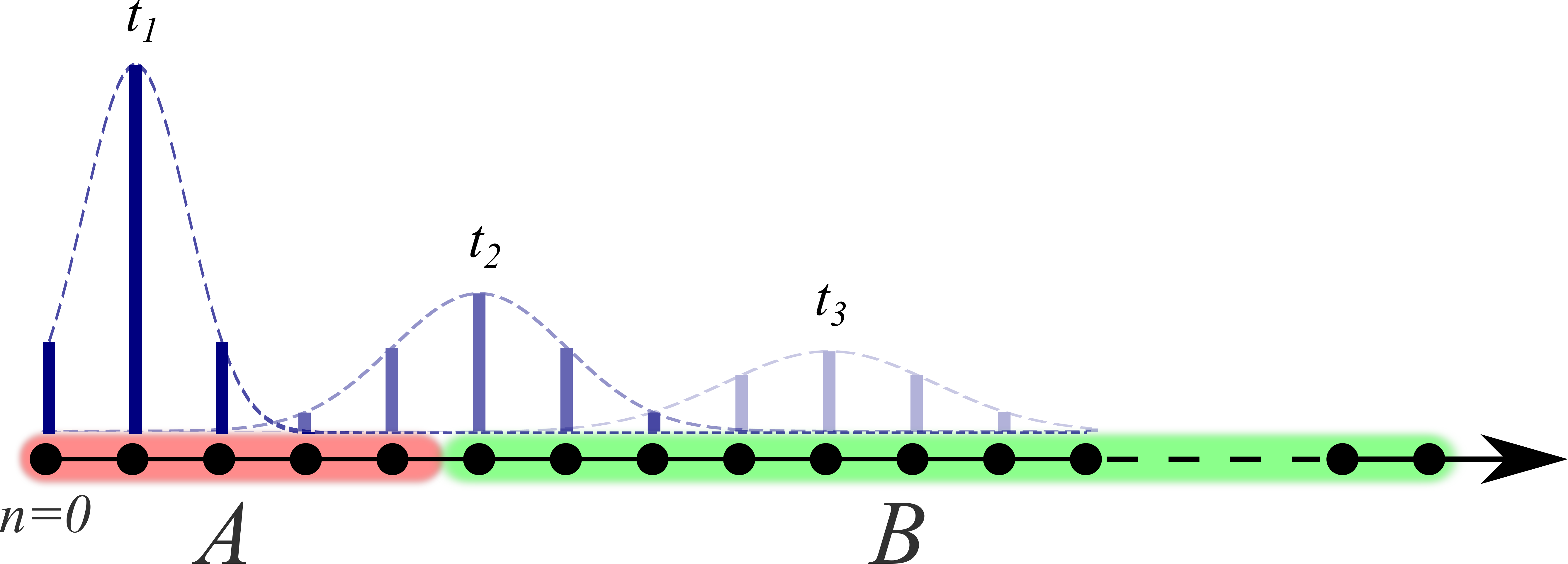}\\
    \vspace{9pt}
    \includegraphics[width=0.6\linewidth]{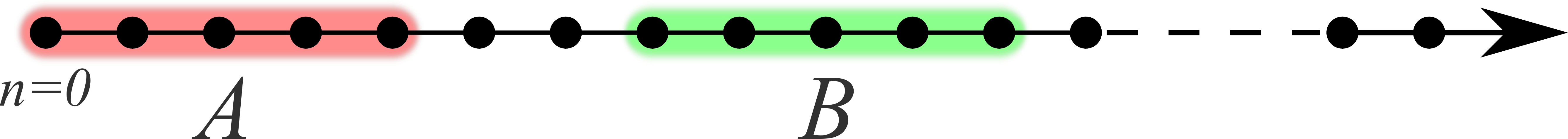}
    \caption{Two possible configurations for subsystems $A$ and $B$ in chord space. Above we show a cut bisection together with a sketch of the Krylov wavefunction propagating in time along the chord states. Here $A$ consists of all chord numbers $n<K$ for a fixed $K$ and $B$ contains the chord numbers $n\geq K$. Below, $A$ and $B$ are disjoint intervals in chord space.}
    \label{fig:subsystems}
\end{figure}

In this work, we explore the density kernel and its associated probes, explicitly using the chord Hilbert space formulation of DSSYK. The chord number basis provides a natural basis for the Hilbert space, which has a similar structure of that of a q-deformed harmonic oscillator. Using this basis, we subsequently explore further quantum information quantities, namely the second moment, the Frobenius norm, the imagitivity and kernel negativity. We calculate these quantities for several subsystems, which we show schematically in figure \ref{fig:subsystems} for the example of a cut bisection and two disjoint intervals. 

The temporal information diagnostics exhibit qualitatively distinct behaviors depending on the chord space partition: growth and saturation for a chord cut, transient growth and decay for finite separated or overlapping intervals, and memory decay for identical intervals. As $\lambda\to 0$, the slope of this growth or decay increases respectively: the generation of temporal quantum information becomes faster as the DSSYK model approaches its semiclassical gravity regime. Since the chord basis coincides with the Krylov basis, this accelerated dynamics admits a natural interpretation in terms of wavefront propagation along the Krylov chain, governed by the position-dependent Lanczos coefficients. Kernel negativity, on the other hand, develops on shorter timescales as $\lambda$ decreases, while its magnitude remains sensitive to the choice of chord space regions.

 Note that the small $\lambda$ limit is associated with two qualitative aspects: the approach to maximal chaos as well as the suppression of quantum gravitational fluctuations and a sharper bulk geometric description. The kernel probes, therefore, are more sensitive to the causal and scrambling dynamics and do not keep a close track of the magnitude of the bulk quantum fluctuations. It is likely that holographic quantum systems are special in this regard, which the kernel probes quantitatively demonstrate.
Let us also elaborate on a different but important distinction about the DSSYK model. The total Hilbert space is a direct sum over the Hilbert spaces with a fixed chord basis. There is no Hilbert space factorization that is natural in the chord basis, and we explicitly use this direct sum structure in our calculations. Consequently, the resulting diagnostics characterize temporal correlations and non-classicality resolved into chord number sectors, rather than conventional entanglement between tensor-factorized subsystems.

Qualitatively speaking, the direct sum structure allows us to consider natural chord number projected sectors, except that these are not similar to symmetry-resolved observables. The latter fact follows trivially, since the Hamiltonian does not commute with the chord number operator. Therefore, under dynamical evolution, different chord number sectors mix non-trivially and our chord number based decomposition becomes similar to a coarse-graining procedure and similar to quantum dephasing channels. This feature is potentially advantageous, since the chord number emerges as a geometric object in the bulk description \cite{Lin:2022rbf,Rabinovici:2023yex}. Conceptually speaking, it hints towards a coarse-graining procedure in which the semiclassical bulk geometric description emerges from the microscopic quantum DSSYK Hamiltonian. This relates to the recent discussion on the interpretation of the semiclassical gravitational path integral as coarse-graining over the fundamentally discrete degrees of freedom  of quantum gravity \cite{Saad:2019lba,Penington:2019kki,Marolf:2020xie,Balasubramanian:2022gmo,Balasubramanian:2022lnw,deBoer:2023vsm,Banerjee:2024fmh,Balasubramanian:2026azk}.

We begin by giving a short review of the spacetime density kernel and its properties, as well as an overview of the double-scaled SYK model and the chord space construction in section~\ref{sec:review}. In the following section~\ref{sec:SDK_for_DSSYK}, we describe our calculation of the spacetime density kernel for the DSSYK model in detail, and reproduce the known results of the DSSYK two-point function and spectral form factor from our approach. Finally, we show our numerical results for the different timelike entanglement measures in section~\ref{sec:numerical_results}. We finally conclude with a discussion on open problems in section~\ref{sec:discussion}. Appendix~\ref{apx:q-hermite} contains our conventions for q-deformed functions, while appendix~\ref{apx:direct_sum} provides additional details on the direct sum structure of the Hilbert space.

\section{Review: Spacetime density kernel and double-scaled SYK}
\label{sec:review}

The conventional notion of entanglement is based on density matrices that are defined on a given Cauchy slice. On the other hand, correlations across time can be recast in terms of a spacetime density kernel which is a generalization of the density operator. This provides a natural description for {\it timelike entanglement}, and its associated measures. The original proposal of timelike entanglement in this context appeared in \cite{Milekhin:2025ycm}, which was subsequently expanded in \cite{Das:2025fcd, Das:2026ifj}. In particular, in \cite{Das:2025fcd}, a higher point generalization of this kernel and its connection with the spectral form factor were also discussed. In this section we briefly review these ideas and the salient features of them.

\subsection{Definition and properties of the spacetime density kernel}
\label{subsec:SDK_def}

Given a factorizable Hilbert space $\mathcal{H}$ and a density matrix $\rho$, time evolution $U_t = e^{-iHt}$,
and a bisection $\mathcal{H} = \mathcal{H}_A \otimes \mathcal{H}_B$\footnote{Note that the subsystems $A$ and $B$ have no constraints. In particular, they can overlap and also individually be the full system.} with matrix unit bases
$\hat{E}^A_{ij} = \ketbra{i}{j} \otimes \mathds{1}_B$ and
$\hat{E}^B_{kl} = \mathds{1}_A \otimes \ketbra{k}{l} $, the definition of the kernel is derived from general correlation functions. For simplicity, let us consider only two-point functions. Given two subsystems $A$ and $B$, respectively defined at two distinct time-slices, $t_1=0$ and $t_2=t$, we consider the kernel operator $\mathcal {T}_{AB} \in \mathcal {B}(\mathcal{H}_A \otimes \mathcal{H}_B)$, where $\mathcal{B}(...)$ is a space of linear operators acting on the corresponding Hilbert space.

The generalized spacetime density kernel (GSDK) $\mathcal{T}_{AB}$ can be understood as the timelike entanglement equivalent to the usual density matrix $\rho$. The kernel encodes the full time dependency of our theory, and for the two-time correlators it is defined such that
\begin{align}
	 \mathrm{Tr} \left[ \mathcal{T}_{AB} (t) (\mathcal{O}_{A} \otimes \mathcal{O}_{B} ) \right] = \mathrm{Tr} \left[ \rho \ \mathcal{O}_{A} U^{\dagger}_{t} \mathcal{O}_{B} U_{t} \right]\,.
	 \label{eq:GSDK_Def}
\end{align}
Furthermore, it is convenient to define the components of $\mathcal{T}_{AB}$, which we call the response tensor $C_{ij;kl} (t)$, for a given choice of basis via
\begin{align}
	\mathcal{T}_{AB}(t) = \sum_{i,j,k,l} C_{ij;kl} (t) E_{ji}^{A} \otimes E_{lk}^{B}\,.
	\label{eq:def_c}
\end{align}
In this way, the two-time correlator from \eqref{eq:GSDK_Def} becomes a simple sum over the operator components 
\begin{align}
    \ev{\mathcal{O}_A \mathcal{O}_B(t)}
    = \text{Tr}\left[\rho\,\mathcal{O}_A\,U_t^\dagger\,\mathcal{O}_B\,U_t\right]
    = \sum_{i,j,k,l} C_{ij;kl}(t)\,\mathcal{O}^{(A)}_{ij}\,\mathcal{O}^{(B)}_{kl}\,,
    \label{eq:correlator_from_SDK}
\end{align}
where $\mathcal{O}^{(A)}_{ij} = \langle i|\mathcal{O}_A|j\rangle$ and $\mathcal{O}^{(B)}_{kl} = \langle k|\mathcal{O}_B|l\rangle$
are the matrix elements in the chosen basis.
We calculate the response tensor as the trace
\begin{align}
    C_{ij;kl}(t) = \text{Tr}\left[\rho\,\hat{E}^A_{ij}\,U_t^\dagger\,\hat{E}^B_{kl}\,U_t\right].
    \label{eq:SDK_def}
\end{align}
The kernel has a normalized trace, ${\rm Tr}(\mathcal {T}_{AB})=1$ and appropriate partial tracing yields a reduced density kernel. Generally, this kernel is non-Hermitian: $\mathcal{T}^\dagger \not = \mathcal{T}$. The adjoint action, in a very precise sense, is sensitive to the time-ordering.\footnote{This can be easily seen at the level of the definition itself:
\begin{eqnarray}
    {\rm Tr} \left[ \mathcal{T}_{AB}^\dagger (\mathcal{O}_A \otimes \mathcal{O}_B)\right]  = {\rm Tr}\left[ \rho \,\mathcal{O}_B(t) \mathcal{O}_A\right] \, ,
\end{eqnarray}
which switches the order of the two local operators, $\mathcal{O}_A$ and $\mathcal{O}_B$.} Given the kernel, it is natural to define density-matrix-inspired diagnostics. In \cite{Milekhin:2025ycm}, several of such probes have been proposed, including:
\begin{eqnarray}
&& {\rm Tsallis \, \, entropy}: \quad    Z_n(t) = {\rm Tr} \left( \mathcal{T}(t)^n\right) \, , \label{eq:moments_kernel}\\
&& {\rm Frobenius \, \, norm}: \quad ||\mathcal{T}(t) ||_2^2 = {\rm Tr} \left( \mathcal{T}(t) \mathcal{T}(t)^\dagger\right) \, , \label{eq:frobenius_norm}\\
&& p-{\rm imagitivity}: \quad \mathcal{I}_p(t) = ||\mathcal{T}(t) - \mathcal{T}(t)^\dagger ||_p  \, , \quad ||M||_p = \left( {\rm Tr}(M^\dagger M)^{p/2}\right)^{1/p} \, . 
\end{eqnarray}
Here $||\ldots ||_p$ denotes the Schatten-norm of the corresponding matrix. The $2$-imagitivity is of particular importance, since it can be shown to upper bound the time-dependent commutator of the corresponding operators. In this sense, the $2$-imagitivity captures a causal influence of the correlation functions. Additionally, the Frobenius norm, Tsallis entropy of order $n=2$, and the $2$-imagitivity are related to each other \cite{Milekhin:2025ycm}. In the following, we will refer to the Tsallis entropy of order $n=2$ as the second moment of the spacetime density kernel.

The kernel encodes both short time-scale physics as well as long time physics. In particular, it has access to spectral correlations, which encode the late-time chaotic behavior of the system as signaled by universal properties of random matrix theory (RMT) \cite{Bohigas:1983er}. The simplest way to see this is to consider a Haar-averaged two-point function of unitary operators, denoted by $\overline{F_2(t)}$. It has been shown that this Haar-averaged two-point function for a thermal density matrix at the inverse temperature $\beta$ can be written as \cite{Das:2025fcd}
\begin{eqnarray}
    \overline{F_2(t)} = \frac{1}{L}{\rm Tr}\left[ \mathcal{T}^{(2)}(t) {\rm SWAP}\right] =\frac{1}{L} \sum_{i,j} C_{ij;ji}^{(\beta)}(t) = \frac{1}{L} \frac{\left| Z\left( \beta/2, t\right)\right|^2}{|Z(\beta)|} \, ,
    \label{eq:Haar_avg}
\end{eqnarray}
where $L$ denotes the dimension of the Hilbert space and SWAP is the standard swap operator. Therefore, the density kernel is related to the spectral form factor (SFF), which is defined in terms of the analytically continued thermal partition function $Z(\beta)$ of the corresponding system. For chaotic quantum systems the SFF has a characteristic slope-dip-ramp-plateau structure, which is a feature of late-time chaos. 

The kernel and its derived quantities above are defined on the basis of a two-point correlation function, which was subsequently generalized for a $(2n)$-point function in \cite{Das:2025fcd}. This makes a direct connection with the physics of scrambling and the GSDK provides us with a common framework in which both early time operator growth physics of out-of-time-ordered correlators (OTOC) and the late time physics of level correlations and spectral chaos are accessible. This connection is explicitly obtained upon Haar-averaging a generic $(2n)$-point function, which yields the $(2n)$-th moment of the spectral form factor. For example, given a $4$-point OTOC the relation between the Haar-averaged OTOC, the corresponding generalized response tensor and the spectral form factor is given by \cite{Das:2025fcd}
\begin{eqnarray}
    \overline{F_4(t)}= \frac{1}{L} \sum_{i,j,l,m} C_{ij, jl, lm, mi}^{(\beta)}(t) = \frac{1}{L} \frac{|Z(\beta/4,t)|^4}{|Z(\beta)|} \, ,
\end{eqnarray}
where the generalized response tensor $C_{ij, jl, lm, mi}$ is a generalization of the response tensor in \eqref{eq:SDK_def}. Again, $L$ is the dimension of the Hilbert space and $\overline{F_4(t)}$ is the Haar-averaged OTOC. In this article, we will not make extensive use of these relations beyond establishing a broader perspective.

It was further shown in \cite{Das:2026ifj} that kernel-derived diagnostics are efficient in capturing the nature of quantum dynamics, and in particular, in the Rosenzweig-Porter model, they can detect how the integrable regime crosses over to a chaotic regime through a multifractal intermediate phase, as the relevant parameter is tuned. Furthermore, in \cite{Das:2026ifj}, a direct connection of the nature of the quantum dynamics with the quantitative measure of non-classicality was also established.  

The amount of non-classicality is quantified by a kernel negativity \cite{Das:2026ifj}, defined as
\begin{eqnarray}
    \mathcal {N}_H = \sum_{\mu_\alpha(t) < 0} |\mu_\alpha(t)| \, , \quad H_{AB}(t) = \frac{\mathcal{T}_{AB} + \mathcal{T}_{AB}^\dagger}{2} \, ,
\end{eqnarray}
where $\{\mu_\alpha\}$ are the eigenvalues of $H_{AB}$. This negativity is upper bounded by the total dimension of the Hilbert space and the purity of the Hermitian kernel, $H$ (see \cite{Das:2026ifj} for more details). As also shown in \cite{Das:2026ifj}, for an RMT-class system, this negativity is bench-marked at approximately half of this bound. Our goal here is to analyze the timelike quantum information of the double-scaled SYK model with this class of diagnostics.

\subsection{Overview of the double-scaled SYK model}
\label{subs:DSSYK_basics}

In the following we introduce the basic properties of the double-scaled SYK (DSSYK) model and its formulation in terms of an auxiliary quantum system described by the chord Hilbert space, which is reviewed in \cite{Berkooz:2024lgq}. In particular, we set up our conventions. The SYK model \cite{SachdevYe:93,Kitaev:2015a} consists of $N$ Majorana fermions $\psi_i$ satisfying the anti-commutation relation $\{\psi_i,\psi_j\}=\delta_{ij}$. The Hamiltonian involves a $p$-body interaction term and is given by 
\begin{align}
	H_{\text{SYK}} = i^{p/2} \sum_{i_1...i_p} J_{i_1 \dots i_p} \psi_{i_1} \dots \psi_{i_p}\,,
	\label{eq:SYK_Hamiltonian}
\end{align}
where the couplings $J_{i_1 \dots i_p} $ are random variables drawn from a Gaussian ensemble with vanishing mean and variance $\ev{(J_{i_1 \dots i_p})^2}=\mathds{J}^2\frac{N}{2p^2}\binom{N}{p}^{-1}$ and we set $\mathds{J}=1$ in the following. This model simplifies in the double-scaling limit \cite{Berkooz:2018qkz,Berkooz:2018jqr}, where we take $N\to\infty$, while the ratio
\begin{align}
	\lambda = \frac{2 p^2}{N}
	\label{eq:lambda}
\end{align}
is kept fixed. Moreover, we define the parameter $q = e^{-\lambda}$. 
The DSSYK model is described by an auxiliary Hilbert space corresponding to the chord Hilbert space $\mathcal{H}$, which is spanned by states $\ket{n}$ of fixed chord number $n=0,1,...$ \cite{Berkooz:2018qkz,Berkooz:2018jqr}. The central goal of this paper is to define the response tensor \eqref{eq:SDK_def} in terms of the chord basis. The action of the Hamiltonian \eqref{eq:SYK_Hamiltonian} on the chord Hilbert space is represented through the transfer matrix 
\begin{equation}
    T=\frac{1}{\sqrt{\lambda}}(a^\dagger+a)\,,
    \label{transfer_matrix}
\end{equation}
where $a^\dagger$ and $a$ are interpreted as creation and annihilation operators. These operators satisfy the algebra of a q-deformed harmonic oscillator
\begin{equation}
    [a,a^\dagger]_q:=aa^\dagger-qa^\dagger a=1
\end{equation}
and their action on a fixed chord state $\ket{n}$ is given by
\begin{equation}
    a^\dagger\ket{n}=\sqrt{[n+1]_q}\ket{n+1}\,,\quad a\ket{n}=\sqrt{[n]_q}\ket{n-1}\,,
    \label{eq:creation_chord}
\end{equation}
where $[n]_q$ is the q-number, which is defined in appendix~\ref{apx:q-hermite} via \eqref{eq:q_number}. Therefore, the chord number operator $\hat{n}$ satisfying $\hat{n}\ket{n}=n\ket{n}$ does not commute with \eqref{transfer_matrix}, which implies that the chord number is not a conserved charge. Moreover, it follows from \eqref{eq:creation_chord} that the transfer matrix \eqref{transfer_matrix} is tridiagonal in the chord basis. This observation leads to equivalence between the chord basis and the Krylov basis with initial state $\ket{0}$ for DSSYK \cite{Rabinovici:2023yex}. 
The dictionary from the physical Hilbert space of the SYK model to the effective chord Hilbert space in the double-scaling limit is to compute ensemble-averaged traces of functions of the Hamiltonian \eqref{eq:SYK_Hamiltonian} via vacuum expectation values of functions of the transfer matrix \eqref{transfer_matrix}. For example, the moments of the Hamiltonian are given by
\begin{align}
    \langle\,\mathrm{Tr}(H^{2k}_\mathrm{SYK})\,\rangle_J = \ev{T^{2k}}{0}\,.
    \label{eq:Relation_SYK_chord}
\end{align}
This can also be seen as a manifestation of the AdS/CFT dictionary, as the transfer matrix $T$ in the triple-scaling limit coincides with the gravitational Hamiltonian of Jackiw--Teitelboim (JT) gravity \cite{Lin:2022rbf,Rabinovici:2023yex}. Unlike in the case of the harmonic oscillator, the Hamiltonian \eqref{transfer_matrix} is not diagonal in the basis $\ket{n}$. Instead, the eigenvectors are parametrized by a continuous variable $\theta \in [0, \pi ]$ such that 
\begin{align}
    T \ket{\theta} = E(\theta) \ket{\theta}\,,
\end{align}
where the eigenvalues are given by \cite{Berkooz:2018qkz}
\begin{align}
    E(\theta) = - \frac{2 \cos(\theta)}{\sqrt{\lambda(1-q)}}\,.
    \label{eq:eigenvalues_T}
\end{align}
We work with conventions such that the chord basis is normalized to $\braket{n}{m}=\delta_{nm}$ and the overlap with the energy basis corresponds to \cite{Berkooz:2018jqr}
\begin{align}
    \braket{n}{\theta} = \frac{H_{n}(\cos(\theta)|q)}{\sqrt{(q;q)_{n}}}\,,
    \label{eq:overlap_chord_energy}
\end{align}
where $H_n(x|q)$ denotes the q-deformed Hermite polynomials defined in appendix~\ref{apx:q-hermite} via the recursion relation \eqref{eq:recursion_q_Herm} and $(q,q)_{n}$ is the q-Pochhammer symbol given by \eqref{eq:pochhammer_finite}. The identity operator is represented in the energy basis as
\begin{align}
    \mathds{1} = \int_{0}^{\pi} d\theta \mu(\theta) \ketbra{\theta}{\theta}\,,
    \label{eq:identity_operator}
\end{align}
where the spectral density 
\begin{align}
    \mu(\theta) = \frac{(q, e^{\pm 2 i \theta};q)_\infty}{2\pi}\,
    \label{eq:energy_density}
\end{align}
accounts for the multiplicity of the eigenvalues. We use the standard notation that $f(\pm \theta):= f(\theta) \cdot f(-\theta)$ and set $(a_1, a_2, \dots, a_n;q)_\infty :=\prod_{k=1}^{n} (a_k; q)_\infty $. Finally, the partition function of the DSSYK model is \cite{Berkooz:2018qkz,Berkooz:2018jqr}
\begin{equation}
    Z(\beta)=\ev{e^{-\beta T}}{0}=\int_{0}^{\pi} d\theta \mu(\theta)\,e^{-\beta E(\theta)}\,,
    \label{eq:DSSYK_partition}
\end{equation}
and the spectral form factor involves an analytic continuation  $Z(\beta)\to Z(\beta+it)$ of this partition function.

The spacetime density kernel is defined such that it reproduces the two-time correlator for generic operators \eqref{eq:GSDK_Def}. For the DSSYK a general operator takes the form
\begin{equation}
    \label{eq:SYK_matter}
    \mathcal{O}=i^{s/2}\sum_{1\leq i_1<...<i_{s}\leq N}K_{i_1...i_{s}}\psi_{i_1}...\psi_{i_{s}}\,,\quad\;\;\frac{s}{p}=\Delta\,,
\end{equation}
where $K_{i_1...i_{s}}$ is a Gaussian random variable that is drawn from a similar distribution as $J_{i_1...i_{p}}$. In the double-scaling limit we take $s\to\infty$, while $2s^2/N$ is finite and the ratio $\Delta$ is interpreted as the scaling dimension of the operator $\mathcal{O}$ \cite{Berkooz:2018jqr}. The two-time correlator of \eqref{eq:SYK_matter} takes the form
\begin{equation}
    \ev{\mathcal{O}(t) \mathcal{O}}=\ev{\Tr(\mathcal{O}e^{iHt}\mathcal{O}e^{-iHt})}_J\,,
    \label{eq:SYK_correlator_def}
\end{equation}
which is evaluated by expanding the time-evolution operators in powers of the Hamiltonian. Therefore, the correlator \eqref{eq:SYK_correlator_def} is given in terms of the moments $\ev{\Tr(\mathcal{O}H^{k_1}\mathcal{O}H^{k_2})}_J$. Similar to the Hamiltonian moments that appear in \eqref{eq:Relation_SYK_chord}, these moments are evaluated in terms of chord diagrams \cite{Berkooz:2018qkz,Berkooz:2018jqr}. In particular, we have to sum over all chord diagrams, with $k_1$ Hamiltonian nodes inserted to the left and $k_2$ to the right of a matter chord. Intersections between two Hamiltonian chords are weighted by a factor of $q$, while intersections between a Hamiltonian and a matter chord are weighted by a different factor $\Tilde{q}=q^\Delta$ \cite{Berkooz:2018qkz,Berkooz:2018jqr}. The resulting two-time correlator is given in terms of the double integral \cite{Berkooz:2018qkz,Berkooz:2018jqr}
\begin{align}
    \ev{\mathcal{O}(t) \mathcal{O}} = \int_0^\pi d\theta d\phi \, \mu(\theta) \mu(\phi)\, e^{it(E(\theta)-E(\phi))} \frac{(q^{2 \Delta};q)_{\infty}}{(q^{\Delta} e^{i(\pm \theta \pm \phi)};q)_\infty}.
    \label{eq:DSSYK_2pt_fct}
\end{align}
A central goal of the following section is to reproduce this two-point function from the spacetime density kernel of DSSYK.

\section{A spacetime density kernel for the DSSYK model}
\label{sec:SDK_for_DSSYK}

With these ingredients we now define a response tensor $C_{ij;kl}$ for double-scaled SYK. Since the SYK model is $(0+1)$-dimensional, the chord Hilbert space $\mathcal{H}$ lacks a factorization $\mathcal{H}=\mathcal{H}_A\otimes\mathcal{H}_B$ into Hilbert spaces associated with spatial subsystems $A$ and $B$. Therefore, we define the response tensor for the special case $A$ and $B$ corresponds to the full system. The spacetime density kernel $\mathcal{T}\in\mathcal{B}(\mathcal{H}\otimes\mathcal{H})$ thus acts on $\mathcal{H}\otimes\mathcal{H}$. For basis elements that appear in the definition of the response tensor \eqref{eq:SDK_def} are therefore given by the full chord basis such that $\hat{E}_{ij}^{A} = \ketbra{i}{j}$ and $\hat{E}_{kl}^{B} = \ketbra{k}{l}$, where in both cases the complement subregion corresponds to $\emptyset$. To study the temporal correlation encoded in $C_{ij;kl}(t)$ we study individual blocks of the full response tensor in section~\ref{sec:numerical_results}.

Recall that the response tensor is defined in terms of the Hilbert space trace \eqref{eq:SDK_def}. It follows from \eqref{eq:Relation_SYK_chord} that after averaging over the random coupling we can exchange the trace over the physical SYK Hilbert space that appears in the definition of the response tensor \eqref{eq:SDK_def} with the vacuum expectation value in the auxiliary chord Hilbert space. In particular, the zero chord state $\ket{0}$ defines a tracial state for the operator algebra generated by the observables \eqref{eq:SYK_matter} and the Hamiltonian \eqref{transfer_matrix}, which acts on the chord Hilbert space \cite{Lin:2022rbf,Xu:2024hoc}. Therefore, the response tensor \eqref{eq:SDK_def} takes the form 
\begin{align}
    C_{ij;kl} = \ev{{\rho \ketbra{i}{j} e^{iT t} \ketbra{k}{l} e^{-iTt}}}{0}\,,
    \label{eq:response_tensor}
\end{align}
where $\hat{E}_{ij}^{A} = \ketbra{i}{j}$ and $\hat{E}_{kl}^{B} = \ketbra{k}{l}$ correspond to the full chord basis. By inserting the identity operator in the energy space representation \eqref{eq:identity_operator} twice into this expression for the response tensor, we find 
\begin{equation}
    \begin{split}
    C_{ij;kl}(t) &=\ev{{0\vert\rho \vert i}}\int d\theta d\phi \,\mu(\theta) \mu(\phi) \braket{j}{\theta}
    e^{i tE(\theta)}  \langle\theta\ketbra{k}{l}\phi\rangle e^{-itE(\phi)}\braket{\phi}{0}\\
    &=\ev{0|\rho|i} \int d\theta d\phi \, \mu(\theta) \mu(\phi) \,
    e^{i t(E(\theta)-E(\phi))} \cdot \frac{H_j(\cos \theta |q) H_k(\cos \theta|q)}{\sqrt{(q;q)_j (q;q)_k}} \ 
    \frac{H_l(\cos \phi |q)}{\sqrt{(q;q)_l}}\,,
    \end{split}
    \label{eq:kernel_integral}
\end{equation}
where in the second equality we inserted the overlap \eqref{eq:overlap_chord_energy} between a chord state and an energy eigenstate and we used that $H_0(x|q) = 1$ and $(q;q)_0 = 1$, which follows from \eqref{eq:recursion_q_Herm} and \eqref{eq:pochhammer_finite}. We have thus obtained an explicit form of the components of the generalized spacetime density kernel, which we can use to calculate timelike entanglement measures introduced in section~\ref{subsec:SDK_def} for the double-scaled SYK model. Before studying the these measure we show that the response tensor \eqref{eq:kernel_integral} satisfies the defining properties that we reviewed in section~\ref{subsec:SDK_def}. In particular, the response tensor has to reproduce the two-point function and the SFF of DSSYK.

The two-time correlation function of generic operators is obtained by summing over the components of the spacetime density kernel  in an appropriate basis \eqref{eq:correlator_from_SDK}. In the explicit case of DSSYK the components of a random operator $\mathcal{O}$ \eqref{eq:SYK_matter} corresponds to a diagonal matrix in the chord space. This becomes evident by the fact that inserting an operator chord on a state with $n$ open chords yields $n$ intersections between operator and Hamiltonian chords, contributing a factor of $q^{\Delta n}$ \cite{Berkooz:2018jqr}. Therefore, we can effectively represent a random operator as a diagonal matrix $\mathcal{O}_{ij} = q^{i\Delta}\delta_{ij}$. The components of this matrix enter \eqref{eq:correlator_from_SDK}, and we choose the zero chord state $\rho_0 = \ketbra{0}$ as a reference state.
This implies
\begin{equation}
    \begin{split}
     \ev{\mathcal{O}(t) \mathcal{O}} &=\sum_{i,j,k,l} C_{ij;kl}(t)\,\mathcal{O}_{ij}\,\mathcal{O}_{kl}= \int d\theta d\phi \, \mu(\theta) \mu(\phi) \,
    e^{i t(E(\theta)-E(\phi))} \\
    & \cdot \sum_{i,j,k,l = 0}^{\infty}\braket{0}{i}\, \frac{H_j(\cos \theta |q) H_k(\cos \theta|q)}{\sqrt{(q;q)_j (q;q)_k}} \ 
    \frac{H_l(\cos\phi |q)}{\sqrt{(q;q)_l}} \ \delta_{ij} \delta_{kl} q^{i\Delta} q^{k \Delta}\,.
    \end{split}
    \label{eq:Sum_kernel}
\end{equation}
It follows from \eqref{eq:recursion_q_Herm} that $H_0(x|q) = 1$ and \eqref{eq:pochhammer_finite} implies $(q;q)_0 = 1$. Therefore, \eqref{eq:Sum_kernel} reduces to a sum over two q-Hermite polynomials
\begin{equation}
    \begin{split}
    \ev{\mathcal{O}(t) \mathcal{O}} = \int d\theta d\phi \, & \mu(\theta) \mu(\phi) \,
    e^{i t(E(\theta)-E(\phi))} \cdot \sum_{k = 0}^{\infty} \frac{H_k(\cos \phi|q) H_k(\cos\theta |q)}{(q;q)_k} \ 
     q^{k \Delta}\,.
    \end{split}
\end{equation}
Using the identity \eqref{eq:n_orth_q_Herm}, we find
\begin{align}
    \sum_{i,j,k,l} C_{ij;kl}(t)\,\mathcal{O}_{ij}\,\mathcal{O}_{kl} = \int_0^\pi d\theta d\phi \, \mu(\theta) \mu(\phi) e^{it(E(\theta)-E(\phi))} \frac{(q^{2 \Delta};q)_{\infty}}{(q^{\Delta} e^{i(\pm \theta \pm \phi)};q)_\infty}\,,
\end{align}
which reproduces the DSSYK two-point function \eqref{eq:DSSYK_2pt_fct} as expected from \eqref{eq:correlator_from_SDK}. 

Additionally, as shown in \cite{Das:2025fcd} and reviewed in section \ref{subsec:SDK_def}, contracting the kernel for a thermal state $\rho_\beta$ with the SWAP operator yields the SFF \eqref{eq:Haar_avg}. Therefore, we calculate the spectral form factor of DSSYK from the generalized spacetime density kernel. Using the equal-spacing finite-temperature regularization of the response tensor, we have
\begin{equation}
    C_{ij;kl}^{(\beta)}(t)=\Tr(y^2E_{ij}y^2U_t^\dagger E_{kl}U_t)=\bra{0}y^2E_{ij}y^2U_t^\dagger E_{kl}U_t\ket{0}\,,
    \label{eq:equal_spacing_regularization}
\end{equation}
where the final expression is a consequence of the fact that $\ket{0}$ is a tracial state for the DSSYK algebra \cite{Lin:2022rbf,Xu:2024hoc}. Here, $y$ is defined such that $y^4=\rho$, and we use the zero chord state $\ket{0}$, which coincides with the infinite-temperature thermofield-double state, for $\rho$ \cite{Lin:2022rbf}. Therefore, we have $y=\ketbra{0}{0}$ and obtain $y^2E_{ij}y^2=\delta_{0i}\delta_{0j}$. Inserting this together with \eqref{eq:equal_spacing_regularization} into \eqref{eq:Haar_avg}, we obtain
\begin{equation}
    \overline{F_{2}(t)}=\frac{1}{L} \lvert \langle 0 \vert U_t \vert 0 \rangle \rvert^2=\frac{1}{L}|\Tr(U_t)|^2=\frac{1}{L}\int d\theta d\phi \,  \mu(\theta) \mu(\phi) \,
    e^{-i t(E(\theta)-E(\phi))}\,,
    \label{eq:SFF_SYK}
\end{equation}
where the representation of $\overline{F_{2}(t)}$ in the energy basis coincides with the absolute value squared of the analytic continuation of the partition function \eqref{eq:DSSYK_partition} at $\beta=0$. Therefore, \eqref{eq:SFF_SYK} indeed reproduces the spectral form factor \eqref{eq:Haar_avg}, as expected from the results of \cite{Das:2025fcd}.

\section{Information flow in DSSYK}
\label{sec:numerical_results}

The spacetime density kernel allows us to study correlations in time, since it encodes the dynamics of different time-slices in one operator. As discussed in Section \ref{sec:review}, we can extract multiple measures, which quantify these correlations, from the spacetime density kernel $\mathcal{T}_{AB}(t)$. In this section we calculate these quantities for the double-scaled SYK model and evaluate them numerically. 

\subsection{Direct sum structure and decoherence}

Before discussing the functional behavior of the kernel measures under time evolution, a few comments are in order. Note that, in the chord basis, the DSSYK Hilbert space admits a natural direct sum structure:
\begin{eqnarray}
    \mathcal{H}=\mathcal H_{< K}\oplus\mathcal H_{\geq K},
\qquad
\mathcal H_{< K}=\bigoplus_{n=0}^{K-1}\mathcal H_n,
\qquad
\mathcal H_{\geq K}=\bigoplus_{n=K}^{\infty}\mathcal H_n\,,
\end{eqnarray}
where $\mathcal{H}_n$ is spanned by the chord state $\ket{n}$. So, unlike a bipartition, the direct sum structure does not contain the conventional notions of entanglement. Rather, it encodes how quantum information is contained in the chord basis. This distinction is important in deriving the density kernel (or, even the usual density matrix) using this chord basis: it yields blocks of the kernel. For example, a low chord number block probes quantum information structure retained below the scale $K$, while its evolution reveals how this information spreads over the entire chord basis. 

Structurally, this is related to the way in which decoherence is realized. This is perhaps best understood in terms of the standard density operator. For example, given the direct sum structure, a generic density matrix can be written as
\begin{eqnarray}
    \mathcal{H} = \mathcal{H}_A \oplus \mathcal{H}_B \quad \implies \quad \rho = \begin{bmatrix}
  \rho_{AA} & \rho_{AB}  \\
  \rho_{AB} & \rho_{BB} 
\end{bmatrix} \, ,
\end{eqnarray}
where $A$ and $B$ define the chord cut partition, and contain the chord number up to $K$ and beyond $K$, respectively. A standard decoherence is a quantum channel that maps this density matrix to a diagonal one
\begin{eqnarray}
    \mathcal{D}(\rho) = \begin{bmatrix}
  \rho_{AA} & 0 \\
  0 & \rho_{BB} 
\end{bmatrix} \, .
\end{eqnarray}
The removal of the off-diagonal terms is understood as the loss of coherence of the corresponding density matrix. To obtain just the $\rho_{AA}$ block, one needs to further perform a projection/post-selection onto the $A$-sector.

Similarly, the density kernel can also be given a chord space block decomposition. The cut, denoted by $K$, can be realized by a projective measurement that distinguishes between the low chord numbers and the high chord numbers\footnote{Note that, this distinction implies a separation between the early time and the late time sector of the DSSYK model, since the chord number is directly related to the time evolution of the respective state \cite{Rabinovici:2023yex}.} and therefore the ``subsystem density kernel'' can be viewed as the projected component of the density kernel, after a coarse-graining/decoherence in the chord space. In brief, this plays the role of a projective quantum channel to the general density kernel. All corresponding kernel derived probes now explore the dynamics of quantum information under such a projective quantum channel.

It is straightforward to make these statements mathematically precise. For example, given a direct sum decomposition of the Hilbert space $\mathcal{H} = \mathcal{H}_A \oplus \mathcal{H}_B$, the density kernel has a similar decomposition: $\mathcal{T}_{AB} = P_A \mathcal{T} P_B$, where $\mathcal{T}$ is the full density kernel and $P_{A,B}$ are the projectors onto the respective chord number blocks. Evidently, $P_A+P_B=1$ and $P_A P_B=\delta_{AB}P_B$. It is now straightforward to check that the Frobenius norm of the full density kernel has the following decomposition
\begin{eqnarray}
    ||\mathcal{T}||_2^2 = ||\mathcal{T}_{AA}||_2^2 + ||\mathcal{T}_{BB}||_2^2 + ||\mathcal{T}_{AB}||_2^2 + ||\mathcal{T}_{BA}||_2^2 \, , 
\end{eqnarray}
which gives an interpretation of chord block resolved temporal information. See appendix \ref{apx:direct_sum} for a proof of this. Subsequently, a dephasing map can be defined as: $\mathcal{D}(\mathcal{T}) = P_A \mathcal{T} P_A + P_B \mathcal{T} P_B$, which simply allows us to rewrite the above expression as: $||\mathcal{T}- \mathcal{D}(\mathcal{T})||_2^2 = ||\mathcal{T}_{AB}||_2^2 + ||\mathcal{T}_{BA}||_2^2$. This gives a natural interpretation of the Frobenius norm of $\mathcal{T}_{AB}$, in terms of the full kernel and the dephased kernel. This structure is similar to what is used in block-decomposed coherence resource theories, see {\it e.g.}~\cite{RevModPhys.75.715}. The chord space direct sum already provides us with a natural block decomposition in which the off-diagonal entries, {\it e.g.}~$\mathcal{T}_{AB}$, quantifies the inter-block temporal coherence.

As introduced in \cite{Das:2026ifj}, there is a natural Hermitian operator that can be derived from the density kernel: $H = (\mathcal{T}+\mathcal{T^\dagger})/2$, and a corresponding negativity as: $\mathcal{N}(H) = (1/2) (||H||_1 - {\rm Tr}(H))$. It is now straightforward to prove that
\begin{eqnarray}
    0 \le \mathcal{N}(H) - \mathcal{N}(\mathcal{D}(H)) \le \frac{1}{2} ||H - \mathcal{D}(H)||_1 \, . \label{negine}
\end{eqnarray}
The details of the proof are provided in appendix~\ref{apx:direct_sum}. This inequality provides a natural separation of the non-classicality (which is quantified by the negativity of the Hermitian kernel) into two pieces: $\mathcal{N}(H) = \mathcal{N}(\mathcal{D}(H))+\delta \mathcal{N}$, where $\delta \mathcal{N}$ is now upper bounded by the RHS of \eqref{negine}. Intuitively, it makes perfect sense that ${\cal N}(H) \ge {\cal N}(D(H))$, since the dephasing channel removes quantum coherence, reducing the non-classicality in the system. Furthermore, $\mathcal{N}(\mathcal{D}(H))$ measures the temporal non-classicality after coherence between the two chord sectors is removed and $\delta \mathcal{N}$ quantifies the temporal non-classicality of inter-sector coherence. What is intriguing is that the temporal non-classicality\footnote{Note that negativity-measured non-classicality can be made rather precise. For example, the Hermitian kernel negativity can be shown to bound the violation of the Leggett-Garg inequality in generic quantum systems \cite{Das:2026pvq}, which provides a concrete and operational meaning to non-classicality.} encoded in $\delta{\cal N}$ is upper bounded by the $L_1$-norm of the difference between the Hermitian kernel and its dephased version. While these relations hint towards the existence of a systematic structure that connects the density kernel with several physical probes and processes, we leave a detailed study for a future work. 

\subsection{Closed form of the kernel measures}
\label{sec:closed_form}
In the following, we derive analytical expressions for the timelike information measures discussed in section \ref{sec:review} for the DSSYK model. For the subsytems $A$ and $B$ we choose arbitrary intervals in the chord Hilbert space. To this end, we consider equation \eqref{eq:response_tensor}, where the matrix element $M_{jk}\equiv\mel{j}{U_t^\dagger}{k}$ for $j\in A$ and $k\in B$ together with the vector $v_l(t)\equiv \mel{l}{U_t}{0}$, determines the response tensor. The \emph{escape probability}, i.e., the probability for the zero chord state to spread into subsystem $B$ is given by 
\begin{equation}
    P_B(t)\equiv\sum_{l\in B}|v_l(t)|^2=\|v(t)\|^2\,,
    \label{eq:escape_probability}
\end{equation}
and the total \emph{transfer weight} from $A$ to $B$ is given by 
\begin{equation}
    W(t)\equiv\sum_{j\in A}\sum_{k\in B}|M_{jk}(t)|^2=\sum_{j\in A}P_{j\rightarrow B}(t), \quad P_{j\rightarrow B}(t)\equiv\sum_{k\in B} \lvert \langle k \vert U_t \vert j \rangle \rvert^2 \,.
    \label{eq:transfer_weight}
\end{equation}
As mentioned in section \ref{sec:review}, the  chord space can also be interpreted as the Krylov space \cite{Rabinovici:2023yex}. Therefore, the probabilities $P_B(t)$ and $P_{j\rightarrow B}(t)$ correspond to probability distributions in Krylov space, which quantify how likely it is for the initial state to evolve up to a certain state on the Krylov chain.
The Frobenius norm of the spacetime density kernel \eqref{eq:frobenius_norm}, which is simply the sum over all squared components, is given in terms of these probabilities as
\begin{equation}
    \|\mathcal{T}_{AB}\|_2^2=\sum_{j\in A}\sum_{k,l\in B}|M_{jk}|^2|v_l|^2=P_B(t)W(t)\,.
    \label{eq:Frobenius_as_probability}
\end{equation}
Therefore, the Frobenius norm of the kernel measures the probability that the initial state $\ket{0}$ and the states that constitute subsystem $A$ propagate into subsystem $B$.
To determine the moments $\Tr\, [\mathcal{T}_{AB}(t)^n]$ of the kernel \eqref{eq:moments_kernel}, we study how the spacetime density kernel acts on a vector $u\otimes v$ with $u\in A$ and $v$ as above. We have
\begin{equation}
    \mathcal{T}_{AB}(u\otimes v)=u_0(Mv)\otimes v\,,
\end{equation}
and since every vector in the image of $\mathcal{T}_{AB}$ is of the form $u\otimes v$, the operator $\mathcal{T}_{AB}^2$ maps all vectors onto multiples of $(Mv)\otimes v$. It is obvious that choosing $u=Mv$ yields an eigenvector with eigenvalue $(Mv)_0=P_B(t)$. Since this is the only non-zero eigenvalue, we obtain for the higher moments of the kernel as
\begin{equation}
    Z_n(t)=\Tr\, [\mathcal{T}_{AB}(t)^n]=P_B(t)^n\,.
    \label{eq:moment_as_probability}
\end{equation}
Thus, these moments encode the escape probability. Moreover, as shown in \cite{Milekhin:2025ycm}, the 2-imagitivity $\mathcal{I}_2(t)^2$ is determined by the Frobenius norm and the second moment via 
\begin{equation}
    \mathcal{I}_2(t)^2=2\|\mathcal{T}_{AB}\|_2^2-2\,\text{Re}\,Z_2(t)=2P_B(t)(W(t)-P_B(t))=2P_B(t)\sum_{j\in A\setminus\{0\}}P_{j\rightarrow B}(t)\,,
    \label{eq:imagitivity_as_probability}
\end{equation}
where in the second equality we used \eqref{eq:Frobenius_as_probability} and \eqref{eq:moment_as_probability}. Note that in our case, $Z_2(t)$ is always real due to equation \eqref{eq:escape_probability} and \eqref{eq:moment_as_probability}.

For the kernel negativity, we do not obtain a simple analytic expression, but the calculation of negativity can be simplified by writing the spacetime density kernel as a product $\mathcal{T}_{AB}=LR^\dagger$, with $L=\mathds{1}_A\otimes v$ and $R=(e_0\otimes\mathds{1}_B)M^\dagger$. Here, $e_0$ is the unit vector of the vacuum and both $L$, $R$ are matrices of dimension $|A||B|\times|A|$. It is straightforward to check that this product reproduces the kernel. We then rewrite the Hermitian part of the kernel as 
\begin{equation}
    H_{AB}=\frac{1}{2}(LR^\dagger+RL^\dagger)=\frac{1}{2}Z\Omega Z^\dagger\,,
    \label{eq:block_decomposition_hermit}
\end{equation}
where $Z=(L\ R)$ and $\Omega = \begin{psmallmatrix} 0 & \mathds{1} \\ \mathds{1} & 0 \end{psmallmatrix}$. The eigenvalues of this matrix are shared with $K\equiv\frac{1}{2}\Omega Z^\dagger Z$, and we determine the block structure of $K$ to be 
\begin{equation}
    K = \frac{1}{2}
    \begin{pmatrix}
        (Mv)\, e_{0}^{T} & MM^{\dagger} \\
        P_{B}\, \mathds{1}_{A} & e_{0}\,(Mv)^{\dagger}
    \end{pmatrix}\,.
\end{equation}
Thus we have reduced the computational effort to a $2|A|\times2|A|$-dimensional eigenvalue problem.

With these expressions at hand, we derive an upper bound for the timelike entanglement quantities. The escape probability is maximal, when the initial state is contained in $B$ and thus $P_B(t)\rightarrow1$. Similarly, the transfer weight is maximized, when all states that constitute subsystem $A$ are contained in subsystem $B$ such that  $P_{j\rightarrow B}(t)\rightarrow1$ implying $W(t)\leq |A|$. With this, we can deduce the upper bound for the Frobenius norm $\|\mathcal{T}_{AB}\|_2^2\leq |A|$ and the second moment $Z_2\leq 1$. When these two measures are maximized, the imagitivity is given by $\mathcal{I}_2^2=2(|A|-1)$. Using the block decomposition of $H_{AB}$ \eqref{eq:block_decomposition_hermit}, we obtain the upper bound $\mathcal{N}\leq\frac{|A|-1}{2}$ for the kernel negativity.

\subsection{Numerical simulation of the kernel measures}

In the following, we first calculate these timelike quantum-information probes for a set of given chord cut bisection and then for several choices of intervals for the DSSYK Hilbert space. Note that in our numerical calculations, we can only consider a finite number of chord states. Here, we have used a maximum chord number of $L=400$. We have verified that the results remain stable when $L$ is increased further.

\begin{figure}
    \centering
    \includegraphics[width=.95\linewidth]{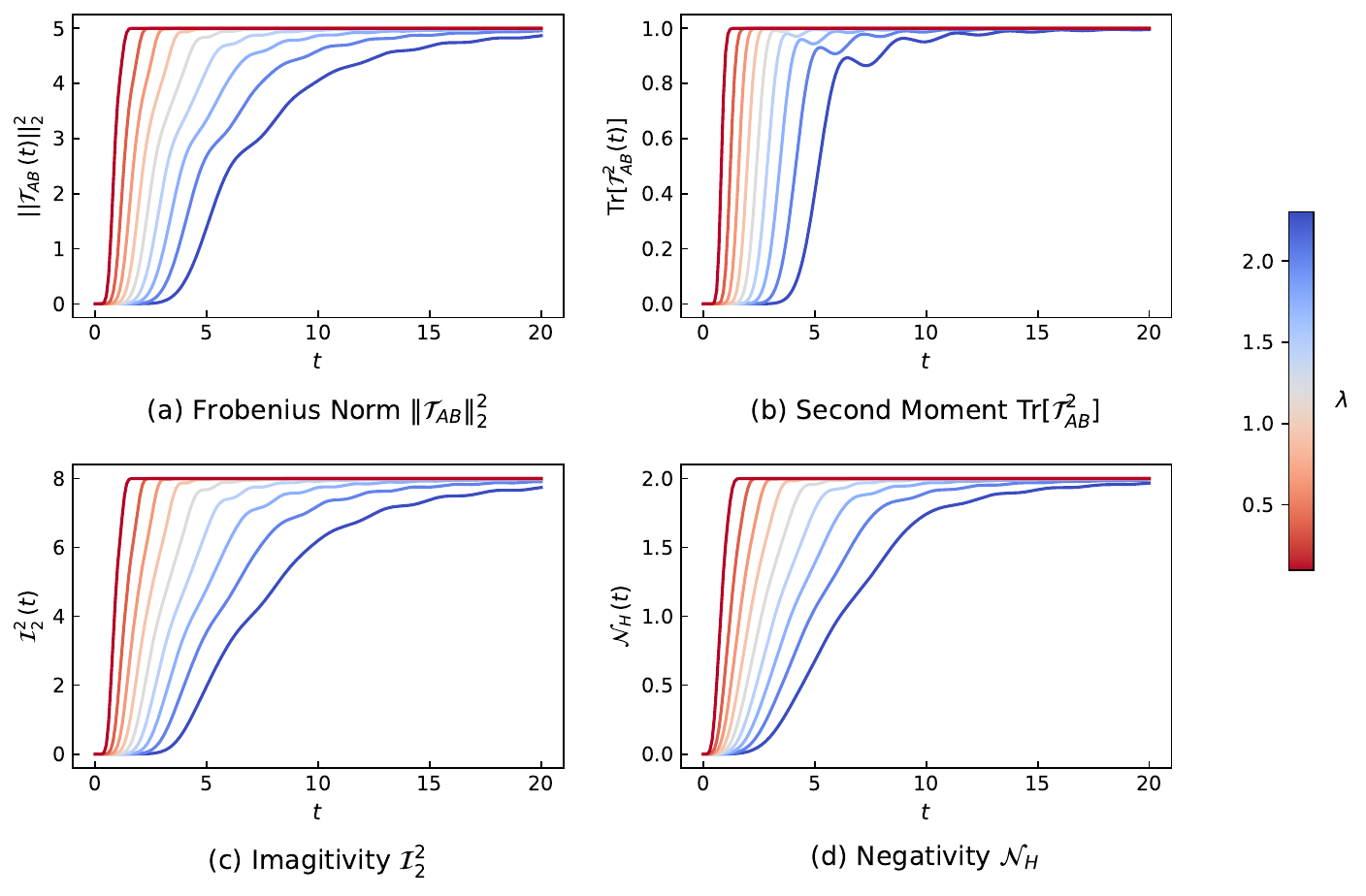}
    {\phantomsubcaption\label{fig:cut_frob}}
    {\phantomsubcaption\label{fig:cut_z2}}
    {\phantomsubcaption\label{fig:cut_img}}
    {\phantomsubcaption\label{fig:cut_neg}}
    \caption{Timelike entanglement measures for several values of $\lambda$ with a cut bisection of the chord space at $K=5$. Subsystem $A$ consists of all chord numbers $n<K$, while $B$ consists of chord numbers $n\geq K$. All measures remain at zero for an onset time, then grow at an increasing rate for small $\lambda$, and eventually plateau at a $\lambda$-independent value that is explained in the main text.}
    \label{fig:cut}
\end{figure}

\paragraph{Chord cut bisection.}~As mentioned above, we divide the chord space by introducing a cutoff parameter $K$, and as subsystems $A$ and $B$ we consider the spaces spanned by basis states $\ket{n}$ with chord numbers $0\leq n<K$ and $n\geq K$, respectively. This choice of subsystems corresponds to the upper panel of figure \ref{fig:subsystems}. We choose the zero chord state $\ket{0}$ as the initial state and a chord cut bisection at $K=5$. The results for the numerical simulation are shown in figure \ref{fig:cut}. The early-time behavior is similar for all measures and all choices of the parameter $\lambda$: the values remain at zero for a short period of time, before starting to increase. This can be explained by the fact, that it takes some time for the state to reach a chord number of $K$, such that correlations to the subsystems at initial time can be observed. Note that we can also interpret the measures in terms of the escape probability \eqref{eq:escape_probability} and the transfer weight \eqref{eq:transfer_weight}. At early times, the escape probability, i.e., the probability for the zero-chord state to evolve into subsystem $B$, is zero, which also determines the Frobenius norm, the second moment and the imagitivity to be zero via equations \cref{eq:Frobenius_as_probability,eq:moment_as_probability,eq:imagitivity_as_probability}. At later times, the behavior of the Frobenius norm, the imagitivity and the Kernel negativity agree qualitatively: we observe a rapid initial growth, followed by a steady growth regime, and eventually the kernel measures reach a plateau value, which is independent of $\lambda$. For the second moment, we do not observe these intermediate regimes, but rather a quick increase to the maximum value, at which it remains. The growth of all the calculated quantities is fastest for a value of $\lambda$ close to zero. In the dual gravity picture, this limit corresponds to the semiclassical regime, which is related to JT gravity in the triple scaling limit \cite{Lin:2022rbf}. In particular, the small $\lambda$ regime corresponds to the most chaotic regime of DSSYK \cite{Maldacena:2015waa}. Therefore, in this regime the information contained in subsystem $A$ propagates the fastest to subsystem $B$.

From our numerical results we can also deduce that the upper bounds derived in section \ref{sec:closed_form} are satisfied for all values of $\lambda$ and the equalities hold at late times. In particular, the Frobenius norm approaches $|A|=5$ at late times, while the second moment approaches 1. The final values for the imagitivity are given by $2(|A|-1)=8$, whereas the negativity becomes $(|A|-1)/2=2$ at late times.

Several comments are in order. First of all, it is intuitively clear why all probes in figure \ref{fig:cut} start from zero and take a while to exhibit growth. Since the initial state is simply the zero chord state, the chord cut bisection is not resolved by the time evolution of this state until a critical time is approached. This sets the scale at which all curves in figure \ref{fig:cut} begin growing dynamically, thus explaining the behavior in the initial regime. 
This regime is followed by a rapidly growing phase for all observables, resulting in a growth of causal temporal correlation and non-classical structure, among other things. The underlying reason is rather universal: quantum information spreads through the chord Hilbert space and becomes accessible across the coarse-graining. This growth is further enhanced in the small $\lambda$-regime where a maximally chaotic semiclassical gravity description emerges \cite{Maldacena:2015waa}.\footnote{Using the example of the Rosenzweig--Porter (RP) model, the authors of \cite{Das:2026ifj} found that the discussed quantities increase most rapidly in the chaotic regime. Except for the second moment, all kernel measures shown in figure~\ref{fig:cut} display the same qualitative behavior as the analogue quantities in the chaotic regime of the RP model \cite{Das:2026ifj}.} It is intriguing that this enhancement happens in the regime when the chord cut bisection becomes resolved and there is flow of quantum information across the cut. Finally, at long enough time-scales, when the finite interval in the chord space associated with $A$ is fully contained within subsystem $B$ the density kernel has no more feature to exhibit and therefore every probe saturates at the upper bound derived above. 

\begin{figure}
    \centering
    \includegraphics[width=0.95\linewidth]{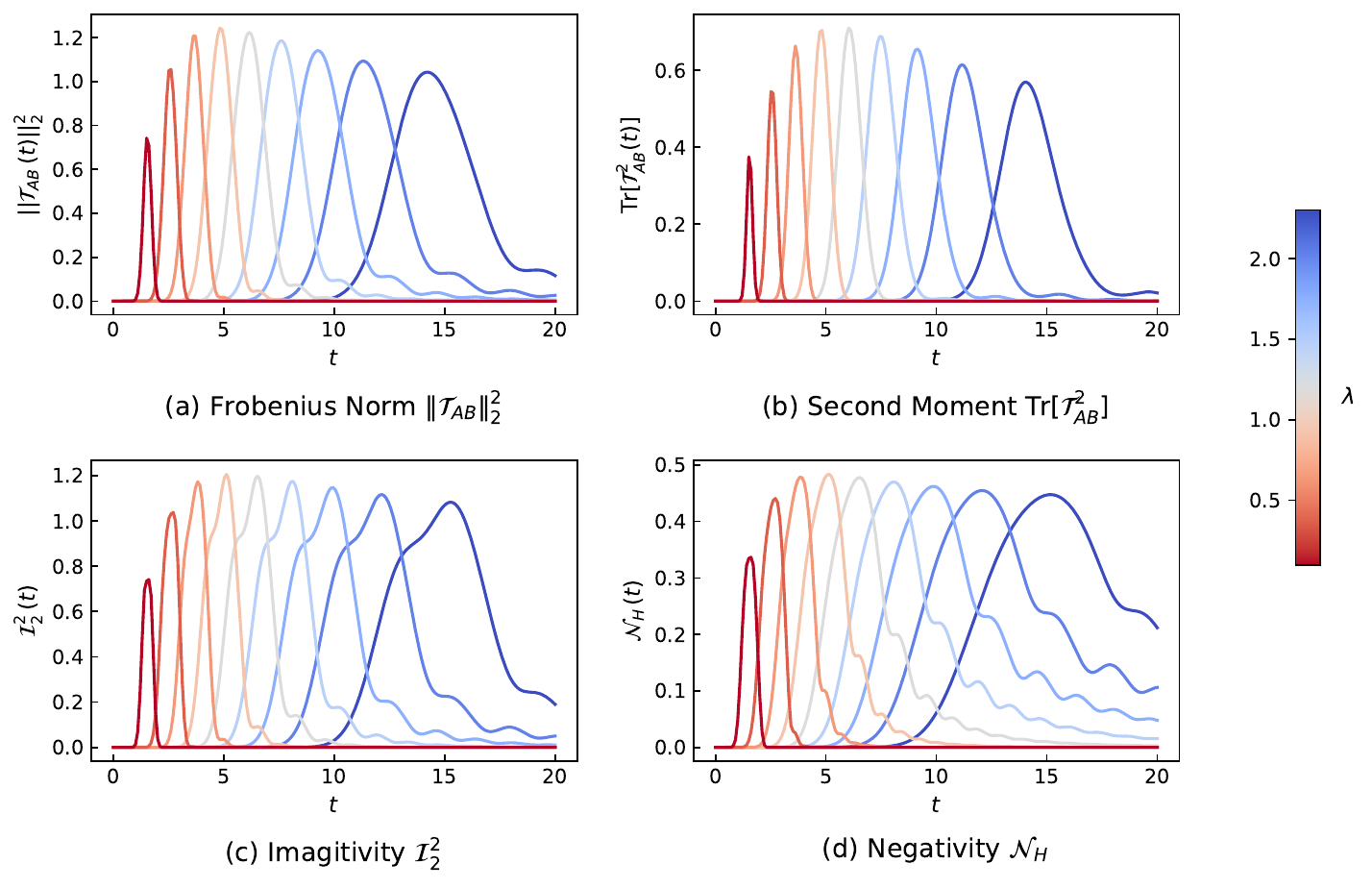}
    {\phantomsubcaption\label{fig:dis_frob}}
    {\phantomsubcaption\label{fig:dis_z2}}
    {\phantomsubcaption\label{fig:dis_img}}
    {\phantomsubcaption\label{fig:dis_neg}}
    \caption{Timelike entanglement measures for two disjoint chord number intervals $A=[0,5]$ and $B=[15,20]$. For all values of $\lambda$, the measures begin to grow after an onset time, then reach a maximum at a time increasing with $\lambda$, before eventually decreasing to zero again.}
    \label{fig:disjoint}
\end{figure}

\begin{figure}
    \centering
    \includegraphics[width=0.95\linewidth]{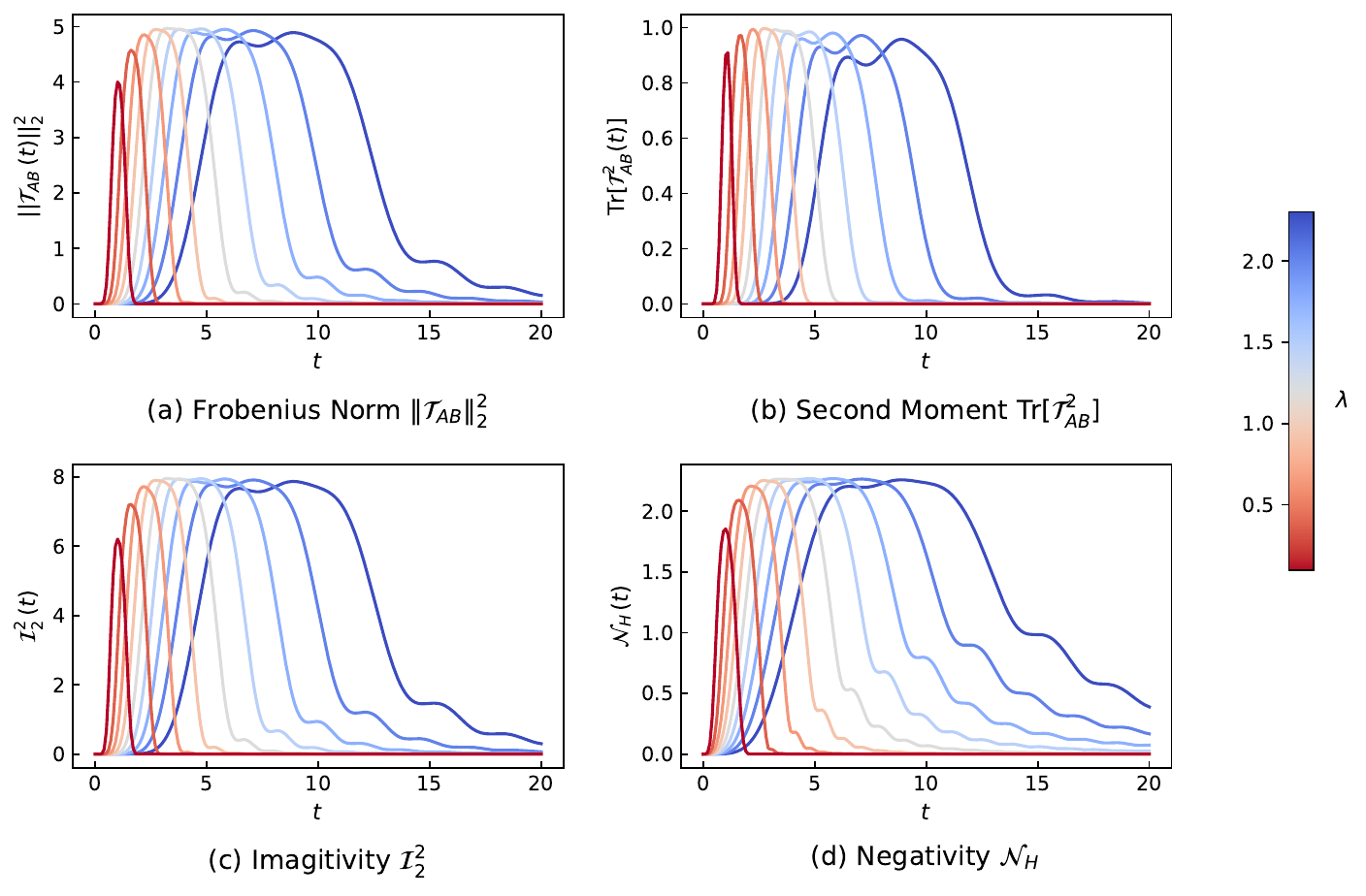}
    {\phantomsubcaption\label{fig:ov_frob}}
    {\phantomsubcaption\label{fig:ov_z2}}
    {\phantomsubcaption\label{fig:ov_img}}
    {\phantomsubcaption\label{fig:ov_neg}}
    \caption{Timelike entanglement measures for two overlapping chord number intervals $A=[0,10]$ and $B=[5,15]$. All measures increase after an onset time, then reach a peak and then decrease to zero. The timescale of this process increases for larger values of $\lambda$.}
    \label{fig:overlap}
\end{figure}

\begin{figure}
    \centering
    \includegraphics[width=0.95\linewidth]{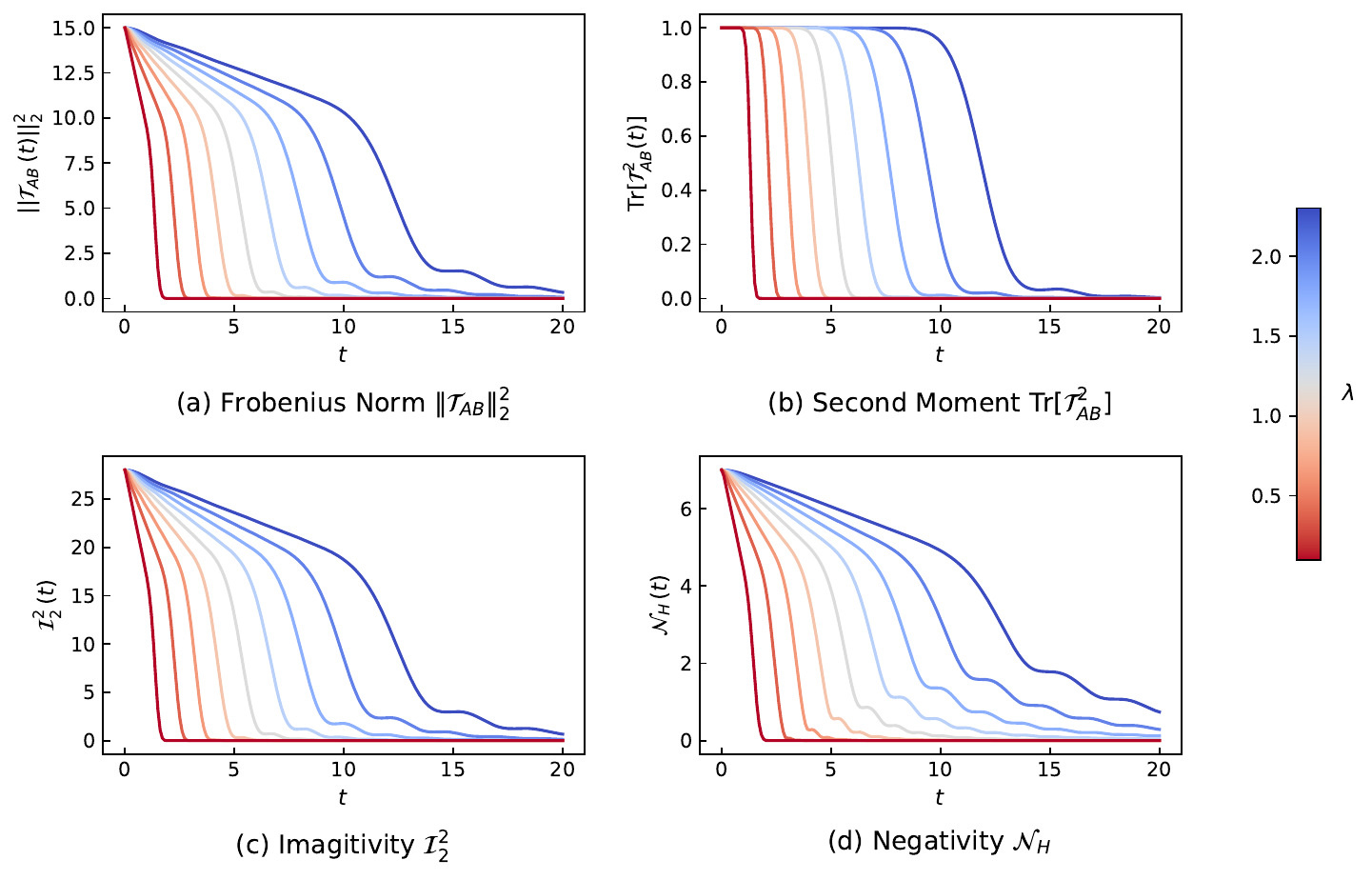}
    {\phantomsubcaption\label{fig:id_frob}}
    {\phantomsubcaption\label{fig:id_z2}}
    {\phantomsubcaption\label{fig:id_img}}
    {\phantomsubcaption\label{fig:id_neg}}
    \caption{Timelike entanglement measures for two identical chord number intervals $A=B=[0,15]$. At the initial time $t=0$, all measures take their maximal values. Subsequently, the measures decrease and eventually settle at zero, while the decrease is faster for smaller values of $\lambda$.}
    \label{fig:identical}
\end{figure}

\paragraph{Interval subsystems of chord space.}~In the following, we calculate the spacetime density kernel and the related timelike entanglement measures for different subsystems $A$ and $B$. We consider two intervals in chord space which are identical, overlapping, or disjoint. The disjoint case is depicted by the lower panel of figure \ref{fig:subsystems}. The results for two disjoint intervals $A=[0,5]$ and $B=[15,20]$ are shown in figure \ref{fig:disjoint}. The results for two overlapping intervals $A=[0,10]$, $B=[5,15]$ are shown in figure \ref{fig:overlap} and for two identical intervals $A=B=[0,15]$ in figure \ref{fig:identical}.

Let us first consider the disjoint subregion case shown in figure \ref{fig:disjoint}. This measures temporal correlations between two {\it distant} subregions, each of which has limited support. The rise-peak-decay behavior has a simple interpretation in this case. Since $A$ and $B$ have no common chord state at initial times, the probes are vanishingly small in this regime. Like in the case of the chord cut bisection, the escape probability \eqref{eq:escape_probability} is zero at early times, which sets the Frobenius norm, the second moment and the imagitivity to zero. As time evolves, common overlaps between the subregions are dynamically created,\footnote{Note that the time-gap of initial response depends on the information probe.} which eventually peak. The peak time-scale is determined by the speed at which information propagates in the chord basis, which is controlled by $\lambda$. In particular, the information propagation is faster the more chaotic the system becomes as $\lambda$ decreases. Eventually, at longer time-scales, since subsystem $A$ eventually propagates out of $B$ the probes, which are given in terms of the probability that $A$ is contained in $B$, also decay. This behavior provides a chord space analogue of the Page curve behavior, although the latter applies to fine-grained entropic measures for a factorized Hilbert space \cite{Almheiri:2019psf,Penington:2019npb,Almheiri:2019qdq,Penington:2019kki}. 
 
The overlapping subregions shown in figure \ref{fig:overlap} bring out a key distinction. Although the rise-peak-decay behavior persists, the interpretation at early-times is slightly different. For an overlapping region, the initial vanishingly small values of the information probes essentially result from the choice of the initial state, which has a non-vanishing support only on the zero chord number eigenstate. As time evolves, the growth and then the subsequent decay can be interpreted in a similar manner as that of the disjoint interval case.

Finally, the identical configuration shown in figure \ref{fig:identical} supports the above interpretation. Note that, in this case, we measure temporal self-correlations. This self-correlation is highest at the initial time, since the regions completely overlap and have access to the entire information content. Again, the kernel measures start at the previously derived upper bound, where the escape probability \eqref{eq:escape_probability} and the transfer weight \eqref{eq:transfer_weight} both take their maximal values. As time evolves, the probes display a finite time memory retention of the chord space subregion. Intriguingly, the Frobenius norm, imagitivity and kernel negativity also display two regimes of approximately linear memory losses, with two different slopes. These two regions are smoothly joined at a time-scale which is reminiscent of the peak time-scale in the earlier two cases. 

To summarize, we have studied three distinct possibilities and three distinct corresponding features. When $A \cap B = \emptyset$, we explore how information transmissions occur between two subregions in chord space. When $A \cap B \not= \emptyset$, we explore the dynamics of transmission and sharing of temporal information. Finally, when $A = B$, we explore the effect of temporal memory retention with a single subregion in chord space. 

\subsection{\texorpdfstring{Qualitative interpretation of $\lambda$-dependency}{Qualitative interpretation of lambda-dependency}}

\begin{figure}[t]
    \centering
    \begin{subfigure}[b]{0.54\textwidth}
        \centering
        \includegraphics[width=\textwidth]{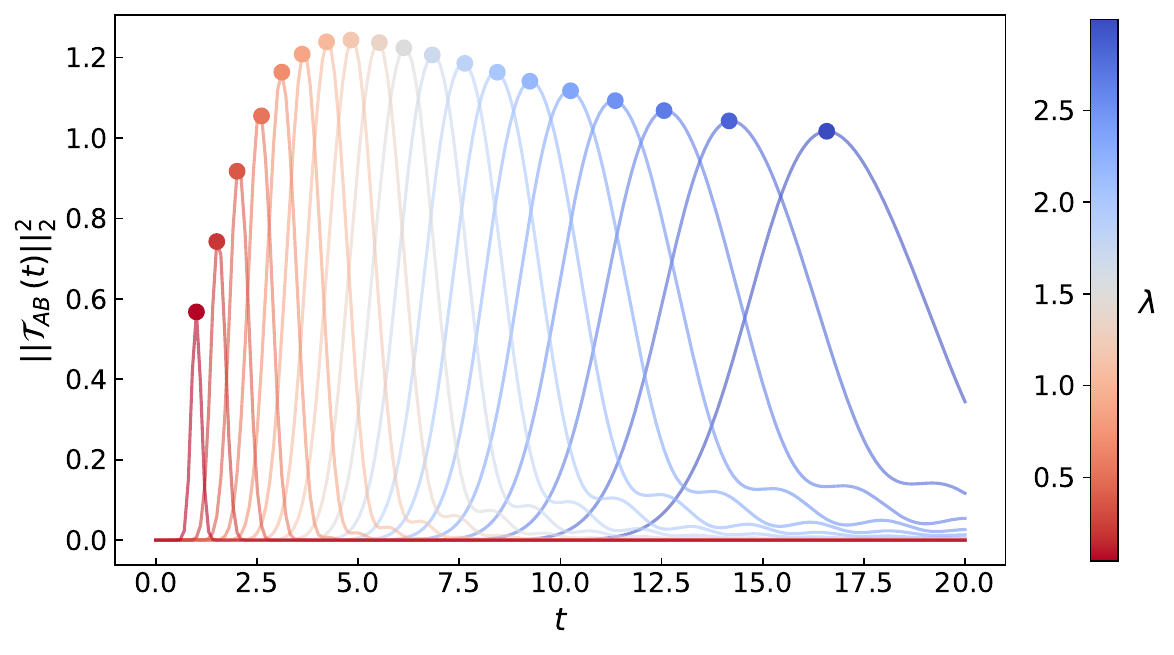}
        \subcaption{}\label{fig:peaks}
    \end{subfigure}
    \hfill
    \begin{subfigure}[b]{0.44\textwidth}
        \centering
        \includegraphics[width=\textwidth]{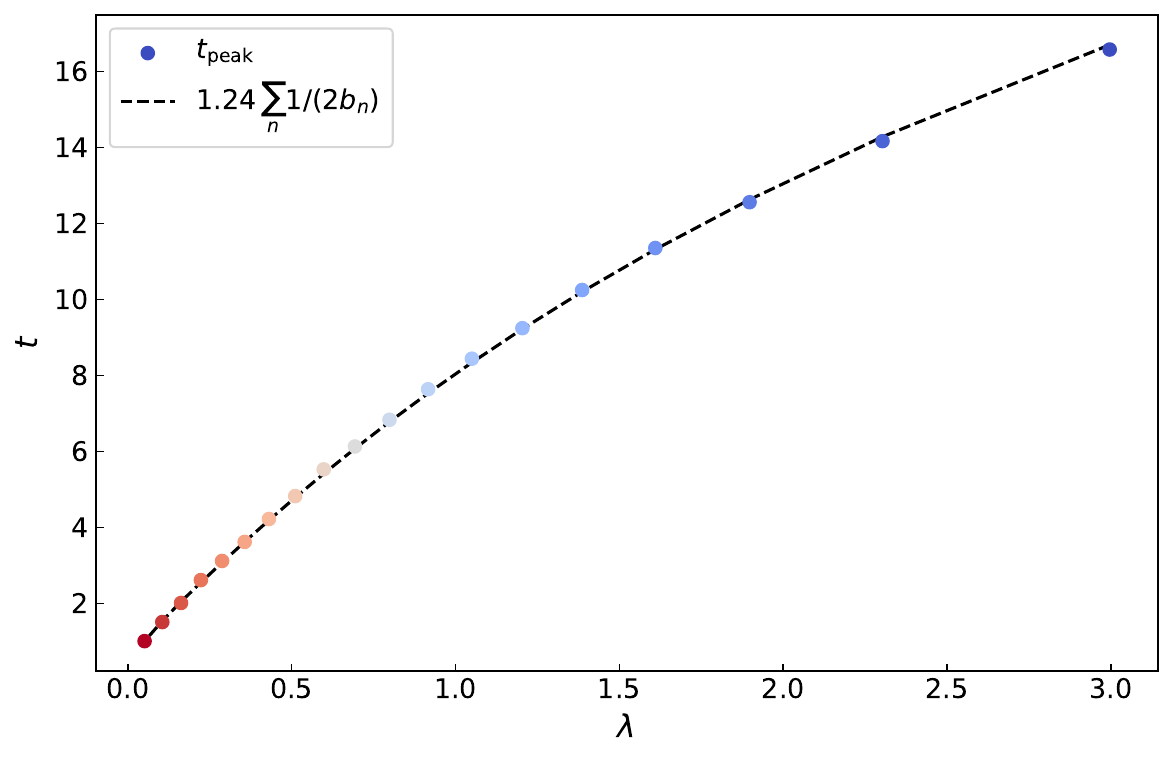}
        \subcaption{}\label{fig:flighttime}
    \end{subfigure}
    \caption{
    (a) Frobenius norm for disjoint intervals $A=[0,5]$ and $B=[15,20]$ in the chord Hilbert space for $19$ logarithmically spaced values of $\lambda$. The simulation was performed for $200$ time-points, the intervals are the same as in figure \ref{fig:disjoint} with a cutoff at $L = 400$. (b) Position of the peaks from figure \ref{fig:peaks}, with a fit of the flight time \eqref{eq:flighttime} in the Krylov chain. The proportionality constant 1.24 is determined by a numerical fit, which matches the peaks in the Frobenius norm very well.}
    \label{fig:peaks_quant}
\end{figure}

In this section, we analyze the dependence of our results on the parameter $\lambda$ further. In particular, we investigate how the position of the peak in the timelike information measures, which is observed for two disjoint subsystems $A=[0,5]$ and $B=[15,20]$, is shifted in time when $\lambda$ changes. For simplicity, we focus on the Frobenius norm, which we show in figure \ref{fig:peaks} for several $\lambda$-values. The height of the peak initially increases with $\lambda$, reaches its maximum in the intermediate regime and decreases for large values of $\lambda$. To understand this behavior further, we can use the interpretation of the chord space as a Krylov space, which has been established in \cite{Lin:2022rbf,Rabinovici:2023yex}. As explained in section~\ref{subs:DSSYK_basics}, due to the tridiagonal form of the transfer matrix $T$, the chord basis precisely coincides with the Krylov basis, and the Lanczos coefficients can be read off from the off-diagonal elements of $T$ via \eqref{transfer_matrix} and \eqref{eq:creation_chord} as $a_n=0$ and
\begin{equation}
    b_n=\frac{1}{\sqrt{\lambda}}\sqrt{[n]_q}\,.
\end{equation}
For our choice of initial state $\rho=\ketbra{0}{0}$, the chord space dynamics can thus be viewed as a hopping problem in Krylov space. In case of a uniform tight-binding chain with coupling $b$, the dispersion relation is given by $E(k)=2b\cos(k)$, which means that waves travel through this chain with a maximal group velocity of $2b$. Using this analogy, we can model the time evolution in chord space as a wave front spreading through Krylov space with a local group velocity of $2b_n$ that depends on the position on the Krylov chain $n$, namely the chord number. For a given chord number $n_B$, we can approximate the \emph{flight time}, which is needed for the initial state to reach $n_B$, by 
\begin{align}
    t_{\mathrm{flight}}(\lambda) \propto \sum_{n=0}^{n_{B}} \frac{1}{2 b_{n}}\,.
    \label{eq:flighttime}
\end{align}
In figure \ref{fig:flighttime}, we plot this time over the parameter $\lambda$, and compare it to the peak times of the Frobenius norm. We obtain a very good match between these two times, confirming our interpretation of the dynamics as a wave front spreading through Krylov space.

\begin{figure}[t]
    \centering
    \includegraphics[width=.65\linewidth]{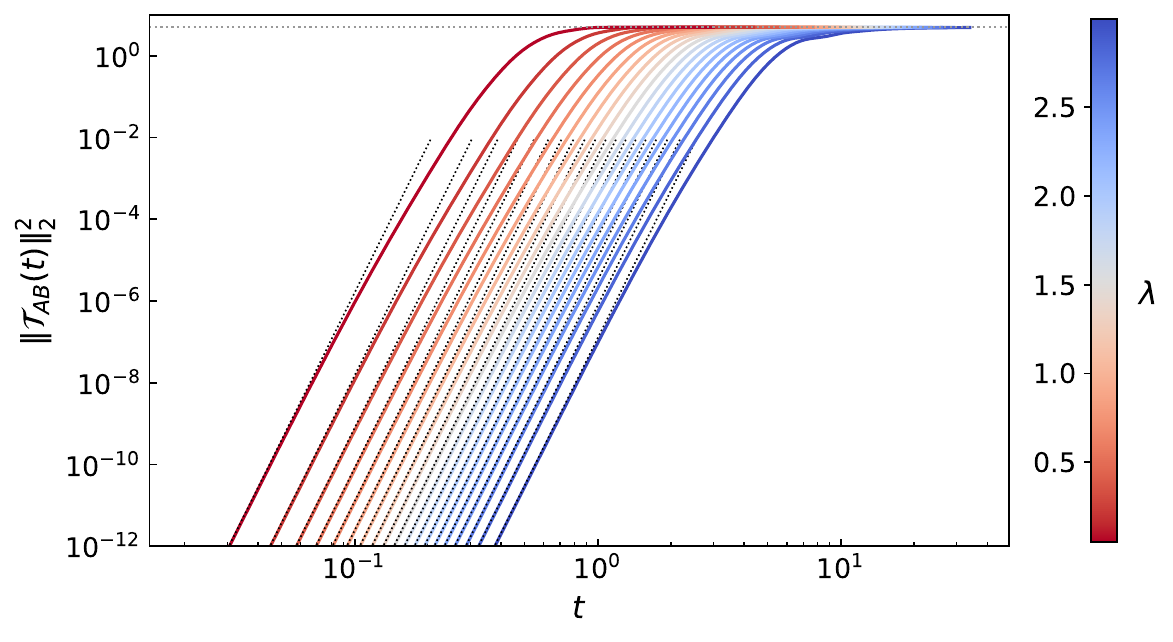}
    \caption{Double logarithmic plot of the Frobenius norm of a spacetime density kernel describing a truncated chord Hilbert space with a cut at $K=5$. For this plot logarithmically spaced values of $\lambda$ were used, to show that the onset of the slope (compare to figure \ref{fig:cut_frob}) rises with $\norm{\mathcal{T}_{AB}(t)}_{2}^{2} \propto t^{2K + 2}$.}
    \label{fig:frob_cut_loglog}
\end{figure}

Additionally, we investigate the early-time behavior of the kernel measures for the example of the Frobenius norm and a cut bisection of chord space. To this end, we consider the matrix elements $\mel{j}{e^{iTt}}{k}$ and $\mel{l}{e^{-iTt}}{0}$, which determine the spacetime density kernel via \eqref{eq:response_tensor} and thus also the Frobenius norm \eqref{eq:frobenius_norm}. Here, we have $j\leq K$ and $k,l>K$. We expand the exponential for early times and obtain that the lowest contribution of the first matrix element is proportional to $t$, since the transfer matrix needs to be applied at least once to obtain a non-zero value. For the second matrix element, the transfer matrix must act at least $K$ times, which means the  lowest contribution is of order $t^K$. In total, the squared Frobenius norm thus scales as $t^{2K+2}$. 
We can confirm this expectation by a numerical calculation, where we use a cut bisection of chord space at $K=5$, and we plot the Frobenius norm over time on a log-log scale, see figure \ref{fig:frob_cut_loglog}. At early times, the Frobenius norm rises linearly in this scale, which indicates a power law dependence. The slope is independent of the parameter $\lambda$, and by a linear fit we can determine its value to be $2K+2$ as expected.

\section{Discussion and outlook}
\label{sec:discussion}

Our results provide the behavior of temporal quantum information in a soluble example with a potentially clean and rich holographic description. Our results further corroborate the usefulness of kernel-derived probes in capturing the nature of quantum dynamics, its non-classicality and its universal framework in general. At the same time, it raises several key questions, especially for the dual geometric description. 

The first and foremost of these aspects is to precisely identify the bulk geometric object that is dual to the density kernel. Since the density kernel is essentially defined in terms of $n$-point correlation functions, such a dual description is guaranteed to exist. In particular, the holographic dictionary between the chord Hilbert space of DSSYK and the fixed length states of JT gravity in the triple-scaling limit \cite{Lin:2022rbf} hints at a precise geometric dual of the density kernel. For example, the density kernel derived from the two-point function must be obtained by massaging the geodesic connecting the two boundary points where the respective operators are inserted. More generally, the density kernel is a non-Hermitian generalization of the density matrix. It is expected that an appropriate generalized gravitational path integral will directly compute this kernel. At this moment, we do not have a precise proposal and it will be important to make a precise identification. 

Our results demonstrate that temporal correlation propagation and memory loss occur on shorter timescales as $\lambda$ decreases towards the semiclassical regime. However, the amplitudes of these diagnostics are not universally enhanced in this limit. This is particularly interesting since it is coupled with the emergence of a semiclassical geometric description. A particularly compelling extension is related to de Sitter holography. The proposed duality in the literature associates the DSSYK time with the observer proper time in the bulk geometry \cite{Narovlansky:2023lfz}. This raises the possibility that the entire density kernel can be associated with the observer's gravitational description.

This question is particularly sharp in the triple-scaling limit at the Schwarzian end of the description, in which a JT gravity-based gravitational description is accessible \cite{Lin:2022rbf}. For example, the ${\rm Tr}(\mathcal{T}^n)$ is qualitatively similar to a Schwarzian bi-local correlator and, therefore, it is given by a JT gravity multi-particle amplitude (or a geodesic network). More generally, at a generic DSSYK point, the density kernel is qualitatively related to an appropriate cyclic gluing of the partition function along a temporal cut. It was already argued \cite{Milekhin:2025ycm} that an analytic continuation of the partition function on an $n$-replicated geometry reproduces the density kernel. Therefore, it is expected that the dual holographic description of the kernel will involve replica wormholes \cite{Penington:2019kki,Almheiri:2019qdq}, which will be very interesting to establish in detail.

More broadly, the direct sum structure of the Hilbert space, its natural association with a dephasing quantum channel between various chord sectors, and the resulting temporal quantum information associated with the density kernel hints towards a unifying framework in which all these come together. The scope of this generalization certainly extends beyond the DSSYK model, and it will be highly interesting to systematically develop this direction into a programme. For example, \cite{PhysRevLett.113.140401,PhysRevLett.118.060502,RevModPhys.75.715} appear to be highly relevant for future developments in these directions.

In summary, the density kernel opens up several aspects for future work: a precise understanding of the bulk, geometric implication of the chord space-based coarse-graining, separating its sensitivity from the semiclassical limit from its sensitivity to quantum chaos, and eventually establishing them in the context of de Sitter holography. We hope to address some of these issues in near future.

\acknowledgments{We would like to thank Rathindra Nath Das, Krishanu Roychowdhury, Nemai C.~Sarkar, Tim Schuhmann and Ashish Shukla for useful discussions. 
J.~E.~and M.~G.~are supported by the Deutsche Forschungsgemeinschaft (DFG) through the Research Training Group GRK 2994, `Particle Physics in the LHC Precision Era'.
J.~E.~and  J.~K.~are supported by Germany’s Excellence Strategy through the Würzburg Dresden Cluster of Excellence ctd.qmat– Complexity, Topology and Dynamics in Quantum Matter (EXC 2147, project-id 390858490)), as well as through the Deutsche Forschungsgemeinschaft (DFG) via the DIP project ‘Holography and the Swampland’ and the German-Korean Project No. ER 301/10 “Emergent spacetime from chaos and entanglement”.  The work of J.~K.~is further supported by the Israel Science Foundation (grant No. 1417/21), by Carole and Marcus Weinstein through the BGU Presidential Faculty Recruitment Fund, by the ISF Center of Excellence for theoretical high-energy physics, by the VATAT Research Hub in the Field of quantum computing, by the ERC Starting Grant dSHologQI (project number 101117338) and by the Kreitman fellowship program at Ben-Gurion University. J.~K.~is grateful to the King's College London for hospitality during the final stages of this work. A.~K.~acknowledges the support of the Humboldt Research Fellowship for Experienced Researchers by the Alexander von Humboldt Foundation and for the hospitality of Theoretical Physics III, Department of Physics and Astronomy, Julius-Maximilians-Universit\"{a}t W\"{u}rzburg and the support from the ICTP through the Associates Programme (2024-2030) during the course of this work. 

\bigskip

\noindent \textbf{\large Statement on AI use}

\noindent Several Anthropic language models were used to assist with calculations throughout this project. In particular Claude Opus 5.5 was used as a coding tool for optimizing the numerical calculations presented in this work. OpenAI GPT-6 model was also used to organize, proofread and finalize the draft. All out-puts were independently reviewed and validated by the authors, who take full responsibility for the accuracy and integrity of this work.}

\appendix

\section{Conventions and properties for q-deformed functions}
\label{apx:q-hermite}

The q-Pochhammer symbol $(a;q)_n$ is defined as 
\begin{align}
    (a;q)_n = \prod_{k=0}^{n-1} \left( 1- a q^k \right)
    \label{eq:pochhammer_finite}
\end{align}
while the q-numbers $[n]_q$ are given by
\begin{align}
    [n]_q:=\frac{1-q^n}{1-q}\,.
    \label{eq:q_number}
\end{align}
Moreover, the q-Hermite polynomials are defined via the recursion relation
\begin{align}
    2x H_n(x|q) = H_{n+1}(x|q) + (1-q^n)H_{n-1}(x|q) \qquad H_{0}(x|q) = 1\,, \quad H_1(x|q) = 2x\,,
    \label{eq:recursion_q_Herm}
\end{align}
with $x=\cos \theta$. They fulfill the orthogonality relation
\begin{align}
    \int_0^\pi d\theta \mu(\theta) H_m(\cos \theta|q)H_n(\cos \theta|q) = (q;q)_n \delta_{nm}\,,
    \label{eq:theta_orth_q_Herm}
\end{align}
where the spectral measure $\mu(\theta)$ is defined in \eqref{eq:energy_density}. Moreover, the polynomials satisfy the identity
\begin{equation}
    \begin{split}
        \sum_{n=0}^{\infty} H_{n}(x|q) H_{n}(y|q) \frac{t^{n}}{(q;q)_{n}} &=
    \frac{(t^{2};q)_{\infty}}{(t e^{i(\theta + \phi)}, t e^{i(\theta - \phi)}, t e^{-i(\theta + \phi)}, t e^{-i(\theta - \phi)}; q)_{\infty}} 
    \label{eq:n_orth_q_Herm} \\
    &=
    \frac{(t^{2};q)_{\infty}}{(t e^{i(\pm \theta \pm \phi)}; q)_{\infty}}\,,
    \end{split}
\end{equation}
with $x=\cos \theta, \, y = \cos \phi$, where we employ the abbreviation $f(\pm \theta):= f(\theta) \cdot f(-\theta)$ and $(a_1, a_2, \dots, a_n;q)_\infty :=\prod_{k=1}^{n} (a_k; q)_\infty $.

\section{Direct Sum Structure of the Hilbert Space}
\label{apx:direct_sum}

Given the direct sum structure, it is natural to introduce a ``coarse-graining" in the chord basis:
\begin{eqnarray}
    {\cal H} = \bigoplus_{k=0}^N {\cal H}_k = {\cal H}_{< K} \oplus {\cal H}_{\ge K} \, , \quad {\cal H}_{< K} = \bigoplus_{k=0}^{K-1} {\cal H}_k \, , \quad {\cal H}_{\ge K}= \bigoplus_{k=K}^N {\cal H}_k \, ,
\end{eqnarray}
where $K$ is a partition that we introduce by hand. We can define a coarse-graining in which ${\cal H}_{\ge K}$ is integrated out (or summed). Subsequently, we can define entropic and information theoretic measures based on this coarse-grained description. 

Let us now use the density kernel to prove a couple of potentially useful identities.\footnote{These proofs were obtained and checked in collaboration with the GPT-6 model.} First of all, under the direct sum structure, the density kernel can be written as
\begin{eqnarray}
   && \mathcal{T} = \mathcal{T}_{AA} +  \mathcal{T}_{BB}+ \mathcal{T}_{AB}  + \mathcal{T}_{BA} \, , \quad \mathcal{T}_{AB} = P_A \mathcal{T} P_B \,, \\
  && P_A P_B = \delta_{AB} P_B \,. 
\end{eqnarray}
The Frobenius norm is given by
\begin{eqnarray}
    || \mathcal{T} ||_2^2 = {\rm Tr} \left( \mathcal{T}^\dagger \mathcal{T}\right) = || \mathcal{T}_{AA}||_2^2 + || \mathcal{T}_{BB}||_2^2 + || \mathcal{T}_{AB}||_2^2 + || \mathcal{T}_{BA}||_2^2 \,. 
\end{eqnarray}
In obtaining the RHS, we have used the defining features of the projection operators, $P_{A,B}$ and the cyclicity of the trace. On the other hand, given the Hermitian part of the kernel
\begin{eqnarray}
    H = \frac{\mathcal{T}+ \mathcal{T}^\dagger}{2} \,, 
\end{eqnarray}
he diagonal dephasing map yields
\begin{eqnarray}
    \mathcal{D}(H) = P_A H P_A + P_B H P_B \,.
\end{eqnarray}
Given a general diagonalizable matrix, $X$, and its spectral data $\{\lambda_i\}$, negativity is defined as
\begin{eqnarray}
    \mathcal{N} (X) = \sum_{\lambda_i<0} |\lambda_i| = \frac{1}{2}\left(||X||_1 - {\rm Tr}(X)\right) \, ,
\end{eqnarray}
where the trace-norm and the trace are given by
\begin{eqnarray}
    ||X||_1 = \sum_{i} |\lambda_i| \, , \quad {\rm Tr}(X) = \sum_i \lambda_i \, .
\end{eqnarray}
Using ${\rm Tr}(\mathcal{D}(H)) = {\rm Tr}(H)$, we therefore obtain
\begin{eqnarray}
    \mathcal{N}(H) - \mathcal{N}(\mathcal{D}(H)) = \frac{1}{2}\left( || H ||_1  - ||\mathcal{D}(H) ||_1\right) \, . \label{NHeq}
\end{eqnarray}
Now, define $V = P_A - P_B$, which gives $V^\dagger = V$ and $V^2 =1$. It is now easy to check that
\begin{eqnarray}
    H + 2 V H V^\dagger = 2 \mathcal{D}(H) \, . 
\end{eqnarray}
Using the triangle inequality, the above relation yields
\begin{eqnarray}
    || \mathcal{D}(H)||_1 \le \frac{1}{2}\left( || H ||_1 + || V H V^\dagger||_1\right)  = || H||_1 \, . 
\end{eqnarray}
Using (\ref{NHeq}) and the above inequality, we immediately obtain 
\begin{eqnarray}
    \mathcal{N}(H) - \mathcal{N}(\mathcal{D}(H)) \ge 0\, . 
\end{eqnarray}
This provides a lower bound for the quantity on the LHS. 

It is also straightforward to prove an upper bound. Towards this, note that the off-diagonal part of the Hermitian kernel is given by
\begin{eqnarray}
    H_{\rm eff-diag} = H - \mathcal{D}(H) \, .
\end{eqnarray}
Using the triangle inequality, we obtain
\begin{eqnarray}
    ||H||_1 - || \mathcal{D}(H)||_1 \le || H_{\rm off-diag} ||_1 \quad \implies \quad \mathcal{N}(H) - \mathcal{N}(\mathcal{D}(H)) \le \frac{1}{2} || H_{\rm off-diag} ||_1 \, . 
\end{eqnarray}
Since the matrix representation of $H_{\rm off-diag}$ can be written as
\begin{eqnarray}
    H_{\rm off-diag} = \begin{bmatrix}
        0 & H_{AB} \\
        H_{AB}^\dagger & 0 
    \end{bmatrix} \quad \implies \quad || H_{\rm off-diag} ||_1 = 2 || H_{AB} ||_1 \, .
\end{eqnarray}
Therefore the upper bound is given by
\begin{eqnarray}
    \mathcal{N}(H) - \mathcal{N}(\mathcal{D}(H)) \le || H_{AB} ||_1 \, . 
\end{eqnarray}
To summarize: we obtain the following upper and lower bound
\begin{eqnarray}
 0  \le \mathcal{N}(H) - \mathcal{N}(\mathcal{D}(H)) \le || H_{AB} ||_1 \, . 
 \label{eq:proof_bounds}
\end{eqnarray}
The lower bound simply says that removing coherence can only decrease negativity or keep it unchanged, which is nicely aligned with the interpretation that negativity quantifies non-classicality of the system. The upper bound appears rather non-trivial. For example, suppose $\mathcal{D}(H)$ is a positive operator, then $\mathcal{N}(H) \le ||H_{AB}||_1$ implies that the trace norm of the Hermitian kernel upper bounds the non-classicality of the system. 

\bibliography{bib.bib}
\bibliographystyle{JHEP}

\end{document}